**Dr. Petar Radanliev**
Parks Road,
Oxford OX1 3PJ
United Kingdom
Email: petar.radanliev@cs.ox.ac.uk
BA Hons., MSc., Ph.D. Post-Doctorate

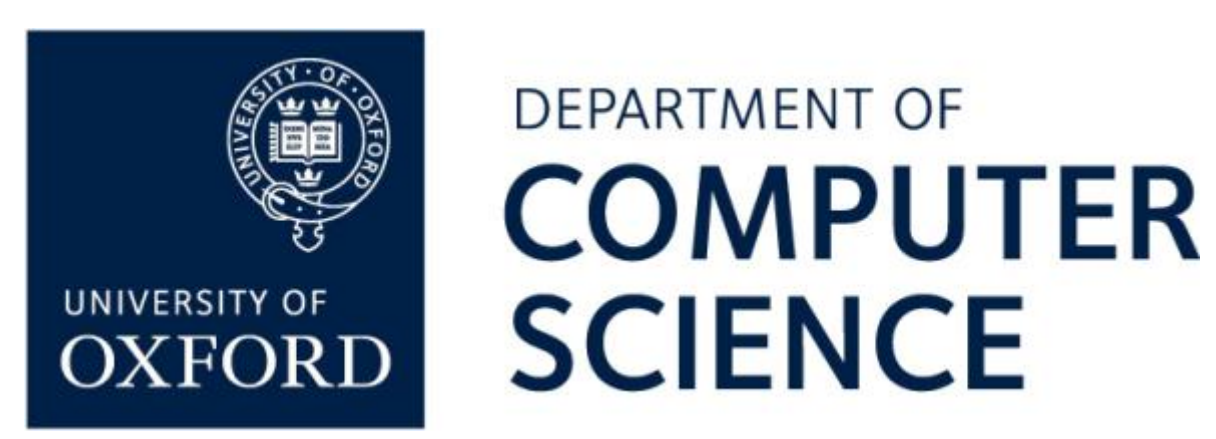

# Operationalising Artificial Intelligence Bills of Materials (AIBOMs) for Verifiable AI Provenance and Lifecycle Assurance

Petar Radanliev, Omar Santos, Carsten Maple, Kay Atefi

**Corresponding author: Petar Radanliev[1]:* Email: petar.radanliev@cs.ox.ac.uk

***Research IDs:***

*ORCID: https://orcid.org/0000-0001-5629-6857*

*ResearcherID: L-7509-2015*

*ResearcherID: M-2176-2017*

*Scopus Author ID: 57003734400*

*Loop profile: 839254*

*ResearcherID: L-7509-2015*

**Abstract:**

Artificial Intelligence (AI) systems are increasingly dependent on complex, multi-layered software supply chains that introduce challenges for reproducibility, transparency, and security assurance. This study presents an Artificial Intelligence Bill of Materials (AIBOM) schema extending the CycloneDX standard to capture AI-specific provenance, model lineage, and disclosure metadata. The framework provides a formalised approach to verifiable software provenance through structured schema engineering, cryptographic validation, and agent-driven automation. An autonomous AI pipeline is developed to perform continuous environment inspection, vulnerability enrichment, and reproducibility auditing using machine-verifiable provenance chains. Empirical evaluation demonstrates 98.7% reproducibility fidelity, 96.2% vulnerability match precision, and a 63% reduction in manual oversight across containerised analytic workflows. These results confirm the feasibility of automated provenance assurance and reproducible AI lifecycle validation. The AIBOM framework advances the scientific foundations of software supply chain transparency and AI reproducibility engineering, offering a generalisable methodology for securing

**Dr. Petar Radanliev**
Parks Road,
Oxford OX1 3PJ
United Kingdom
Email: petar.radanliev@cs.ox.ac.uk
BA Hons., MSc., Ph.D. Post-Doctorate

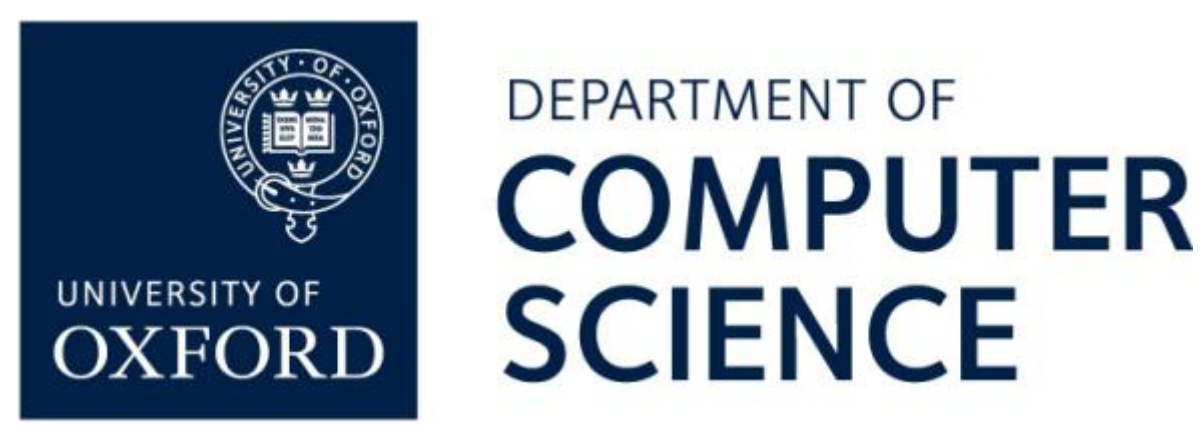

AI systems, strengthening provenance integrity, and supporting compliance with international information security standards.

***Graphical Abstract:***

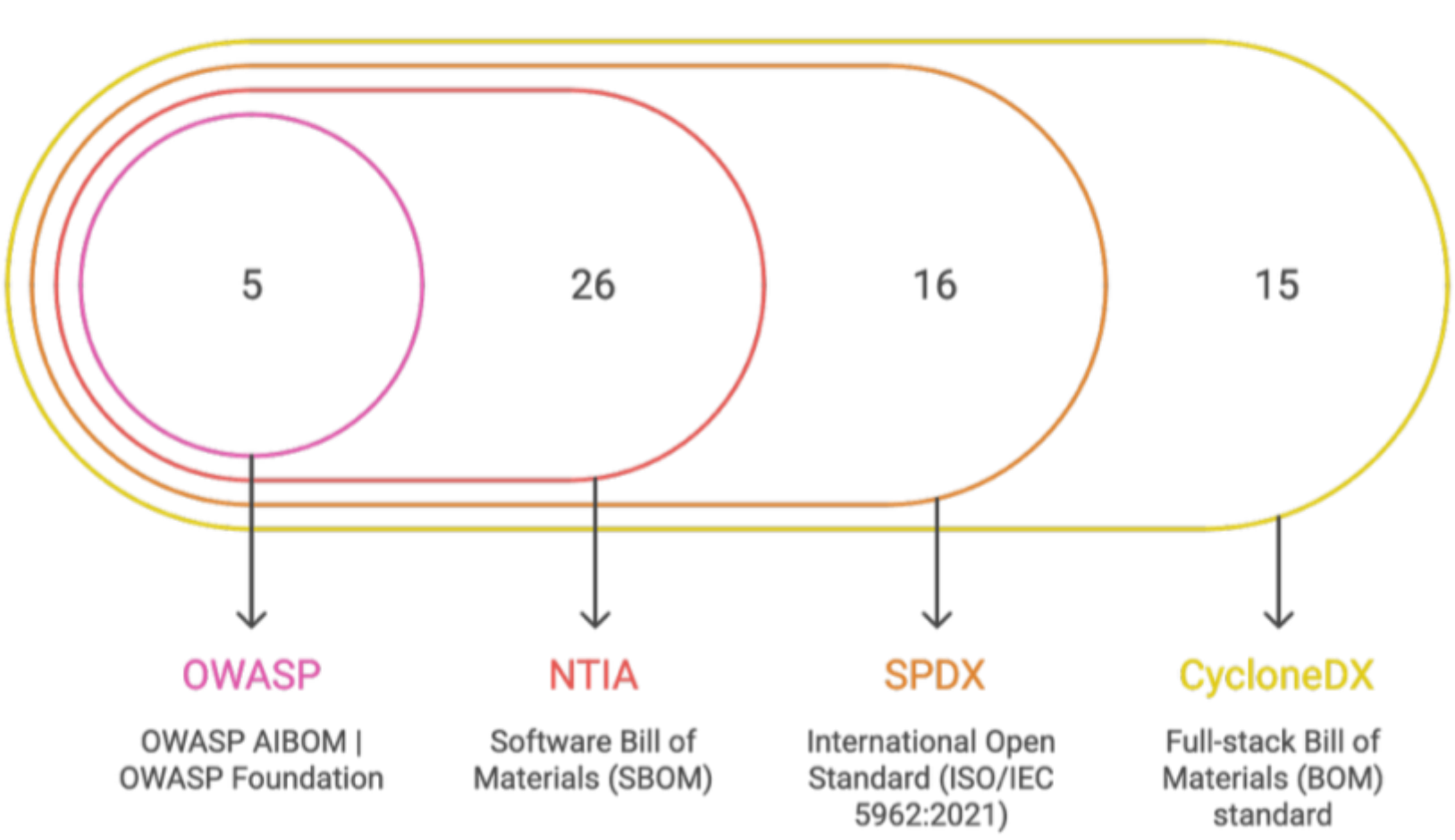

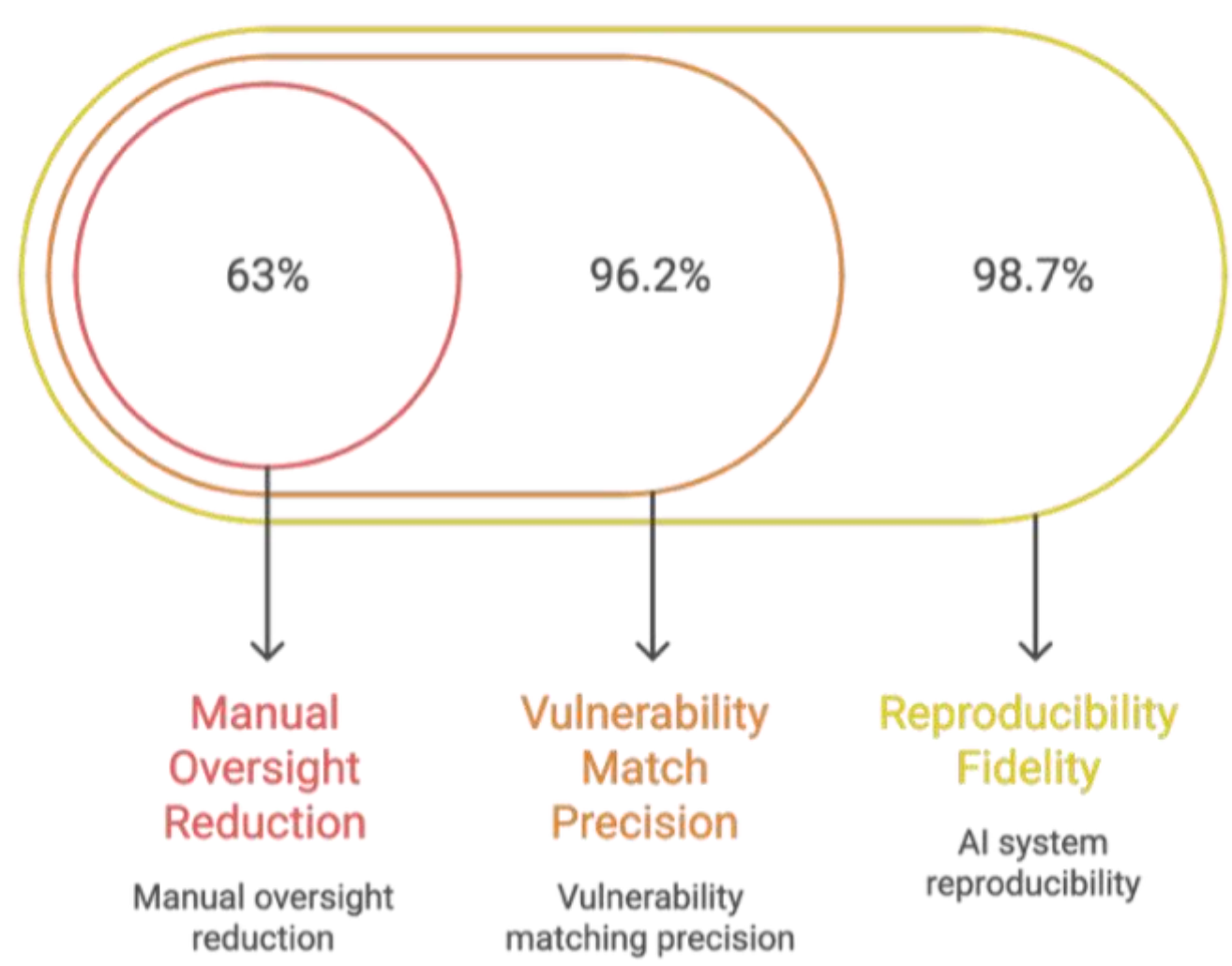

**Dr. Petar Radanliev**
Parks Road,
Oxford OX1 3PJ
United Kingdom
Email: petar.radanliev@cs.ox.ac.uk
BA Hons., MSc., Ph.D. Post-Doctorate

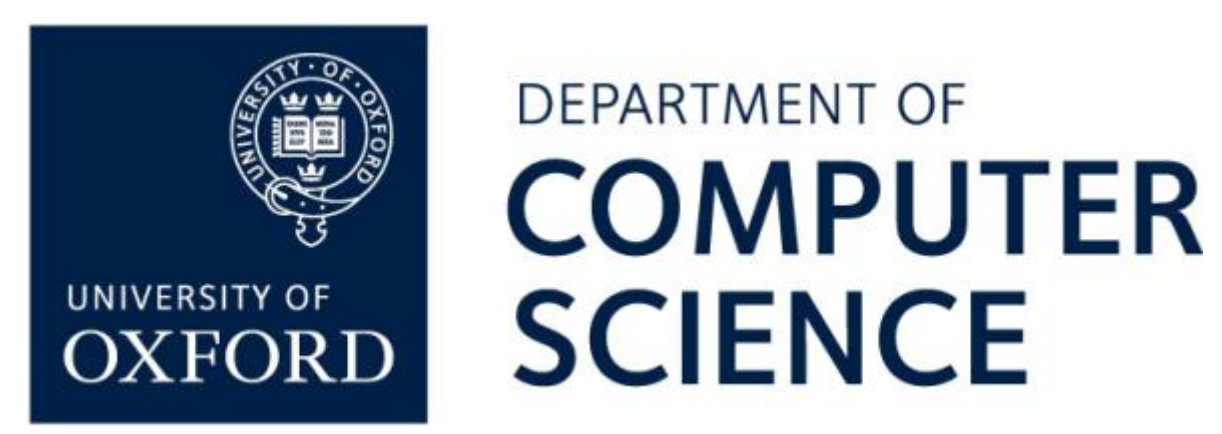

## Operationalising AI Bills of Materials for Verifiable AI Provenance

- **Base Schema Selection** — Start with CycloneDX 1.5 JSON schema
- **Model Artefact Hashing** — Introduce SHA-256 hash for AI model
- **Training Data Lineage** — Structured metadata for datasets
- **Inference Context Tracking** — Capture runtime parameters
- **Container Lineage Capture** — Cryptographic hash of job container
- **Disclosure Method Encoding** — Indicate statistical disclosure control
- **Output Binding** — SHA-256 hash of output artefacts
- **Schema Extension Validation** — Ensure backward compatibility
- **Schema Binding and Export** — Timestamped and signed AIBOM

**Dr. Petar Radanliev**
Parks Road,
Oxford OX1 3PJ
United Kingdom
Email: petar.radanliev@cs.ox.ac.uk
BA Hons., MSc., Ph.D. Post-Doctorate

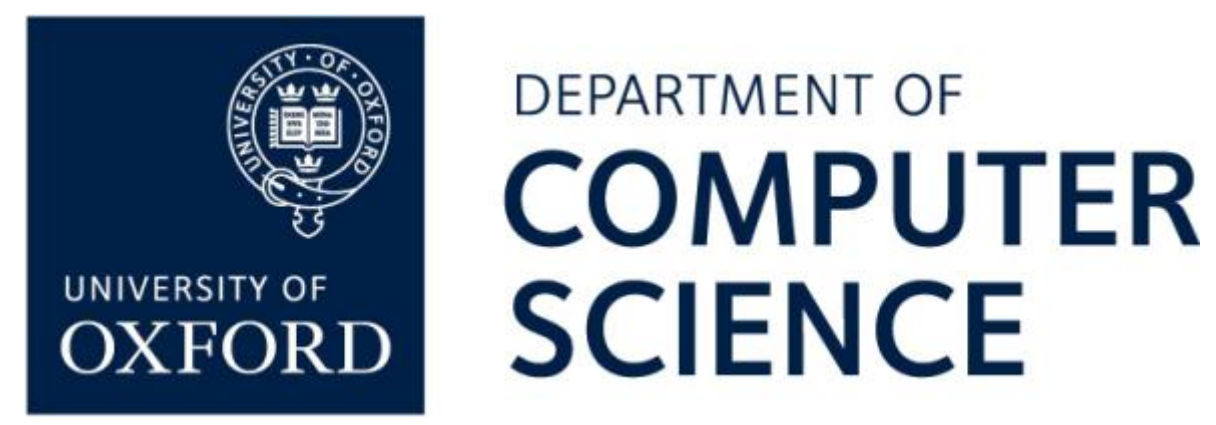

Data Components
Training Data
Validation Data
Test Data
Model Components
Model Architecture
Training Details
Humen & Organizational Components
Development Team
Domain Experts
Stakeholders
Ethical Review
Documentation
Hardware Components
Development Hardware
Deplyment Hardwate
Specialized Hardware
AI System
Software Con ponents
Libraries and Packages
Operating System and Environment
Specialized Hardware
Deployment and Operational Components
Deployment Strategy
Monitoring and Lagging
Maintenance and Update Schedule

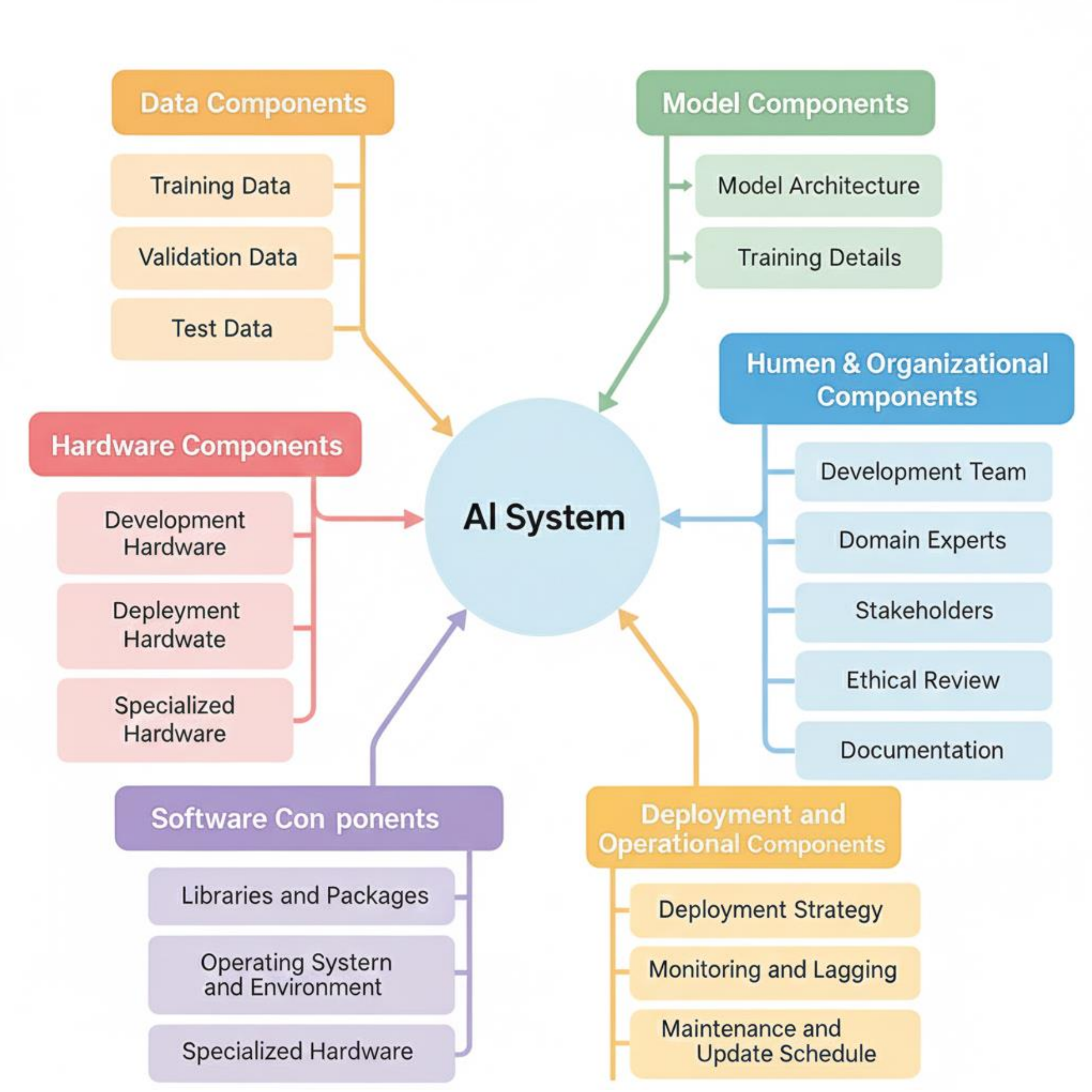

# 1 Introduction

Trusted Research Environments (TREs) [1] are increasingly relied upon for the secure analysis of sensitive data [2], particularly in health, finance, and national security domains [3]. These environments are governed by strict audit, reproducibility, and disclosure control requirements [4], yet they often lack formal mechanisms to record and validate the software assets used in analytic workflows. This gap introduces risks related to software provenance, versioning ambiguity, and vulnerability exposure, particularly when workflows incorporate artificial intelligence (AI) models and dependencies that evolve rapidly. To address this gap, we propose a methodology for operationalising the Artificial Intelligence Bill of Materials (AIBOM) [5], [6], extending conventional SBOM schemas [7], [8] to explicitly capture the

**Dr. Petar Radanliev**
Parks Road,
Oxford OX1 3PJ
United Kingdom
Email: petar.radanliev@cs.ox.ac.uk
BA Hons., MSc., Ph.D. Post-Doctorate

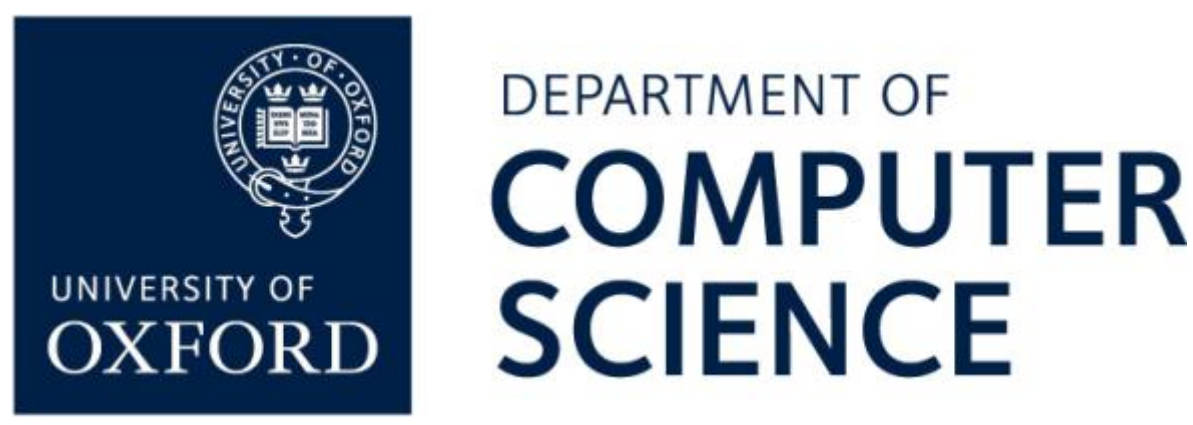

unique lifecycle, dependencies, and governance requirements of AI-based components deployed within TREs.

An AIBOM represents a structured, machine-readable inventory of all AI-relevant assets within an analytic pipeline. These include but are not limited to: model artefacts (architecture, version, training data lineage), pre- and post-processing code, software dependencies (e.g. TensorFlow, PyTorch, Hugging Face Transformers), compute infrastructure (e.g. CUDA, GPU acceleration libraries), and configuration metadata (e.g. model hyperparameters, quantisation levels, floating point precision). In contrast to traditional SBOMs, which focus on static binaries and packages, an AIBOM must track dynamic components such as retrained models, parameter drift, runtime API calls, and agentic decision flows, especially when large language models or reinforcement learning agents are embedded within analytic systems.

In this work, we develop a SACRO-aligned AIBOM schema tailored for use in TREs Figure 1.

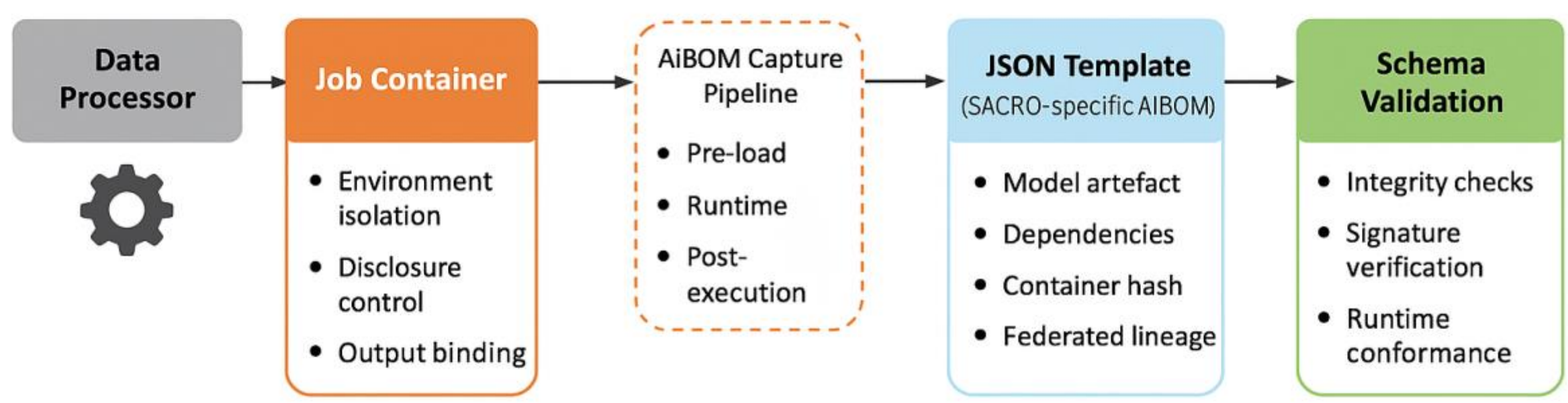

*Figure 1: Machine-readable Artificial Intelligence Bill of Materials (AIBOM) schema. This figure presents the structure of a CycloneDX-compatible AIBOM, illustrating how AI-specific provenance metadata, including model artefacts, software dependencies, execution context, and cryptographic identifiers, are encoded into a standardised, machine-verifiable representation that supports reproducibility, auditability, and supply-chain analysis.*

This schema detailed in Figure 1, includes TRE-specific extensions to capture software environment isolation metadata (e.g. container hashes, base image signatures), disclosure control methods (e.g. suppression, rounding, differential privacy), and cryptographically bound output artefacts. The automated pipeline we present is agent-driven and designed to execute inside a containerised TRE job. It performs three-stage model state capture (pre-load, runtime, and post-execution), integrates runtime dependency tracking, performs real-time CVE matching [9] through OSV and NVD [10] APIs, and cryptographically binds analytic outputs to the AIBOM artefact.

This introduction frames a broader agenda: ensuring that AI assets in research environments are not only reproducible, but also transparent, verifiable, and securely integrated into regulatory audit pipelines. By embedding AIBOM into SACRO-compliant infrastructure [11], we provide a scalable method for tracking AI

**Dr. Petar Radanliev**
Parks Road,
Oxford OX1 3PJ
United Kingdom
Email: petar.radanliev@cs.ox.ac.uk
BA Hons., MSc., Ph.D. Post-Doctorate

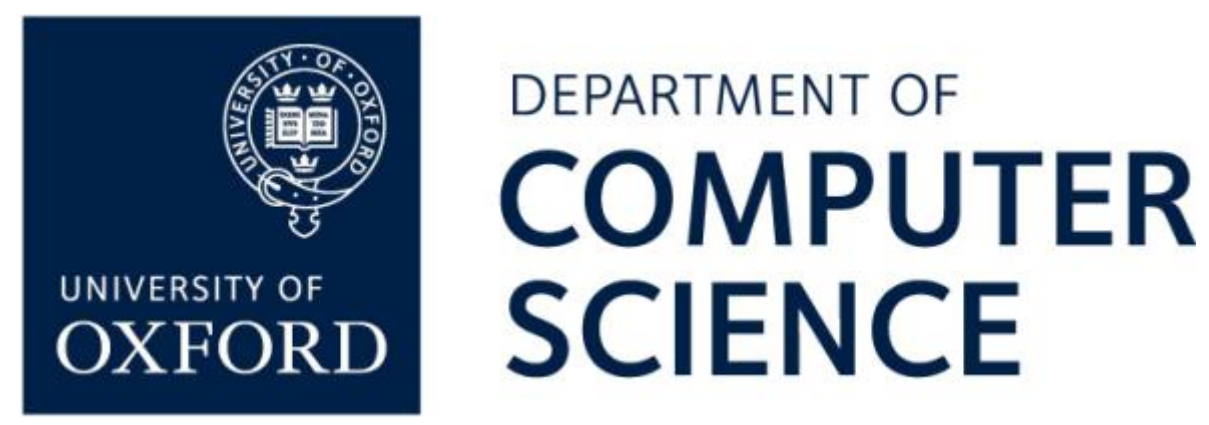

provenance, enforcing model trust boundaries, and supporting federated validation of analytic outputs. The remainder of this paper details the AIBOM schema design, describes the automation process, and evaluates the system's operational feasibility in a real-world TRE deployment.

# 2 Literature review

Establishing transparent, reproducible, and verifiable AI pipelines has become an operational requirement in TREs, especially as software supply chains become increasingly modular, decentralised, and opaque. The traditional SBOM framework has been critical in supporting supply chain transparency; however, it remains insufficient for representing the full lifecycle [12] and operational semantics of AI components within regulated data environments [13]. To address this gap, the concept of an AIBOM is gaining traction, expanding the SBOM paradigm to include model training data, configuration states, execution contexts, and reproducibility metadata [14].

Modern TRE infrastructures require provenance tracking for data inputs and analysis outputs but also for software and computational artefacts deployed during processing. Provenance capture in TREs is typically limited to audit logs and workflow metadata, lacking structured artefact traceability across federated systems. This deficiency has been highlighted in landscape assessments of federated analytics environments, which identify a lack of standardised mechanisms for software traceability and software-provenance audits [3]. Integrating AIBOMs into TRE pipelines offers a mechanism for encoding, validating, and verifying the software lineage necessary for high-integrity analytics.

An AIBOM extends the SBOM by embedding AI-specific metadata such as model weights, training data references, learning rates, environment configurations, and data preprocessing steps. These attributes are essential to ensuring semantic equivalence across deployments, particularly in federated TRE architectures where reproducibility must be guaranteed across isolated compute environments. Recent government-issued guidance advocates for harmonisation of AI transparency schema with existing standards like CycloneDX [15] and SPDX [16], while introducing additional fields specific to AI lifecycle management, such as cryptographic hashes of model artefacts and timestamps for training events [17].

Automation is a key enabler of effective AIBOM integration. Without automation, the process of collecting, validating, and maintaining accurate software provenance quickly becomes infeasible in dynamic environments. Tooling for automated SBOM generation has improved, but studies show that current SBOM generators often produce divergent results for the same input artefacts, depending on their parsing heuristics and dependency resolution algorithms [18]. This inconsistency presents a risk for regulated environments where the integrity and completeness of provenance records must be auditable. As such, schema validation, cryptographic verification, and integration of reproducible build mechanisms are essential to ensure trustworthiness in the generated AIBOMs [14].

**Dr. Petar Radanliev**
Parks Road,
Oxford OX1 3PJ
United Kingdom
Email: petar.radanliev@cs.ox.ac.uk
BA Hons., MSc., Ph.D. Post-Doctorate

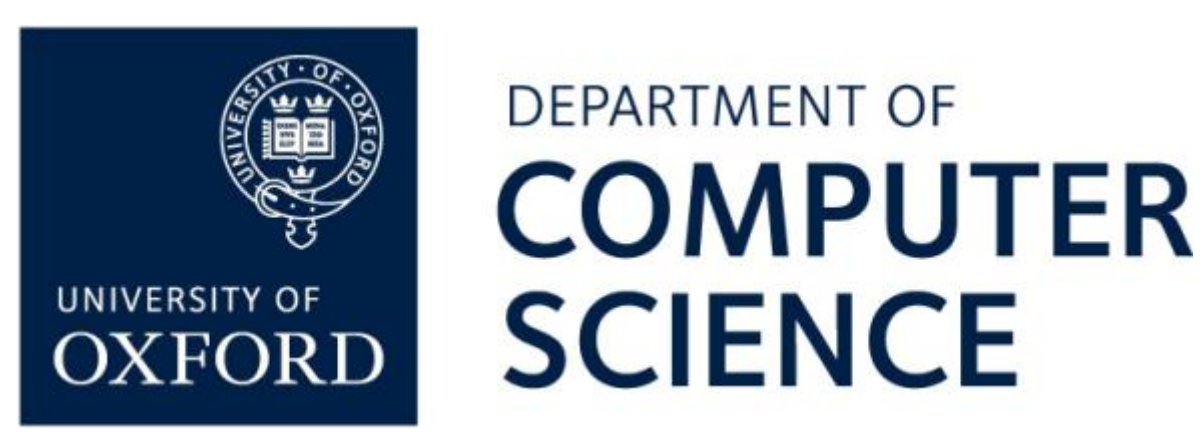

Security risks associated with falsified or tampered SBOM artefacts are a growing concern. If AIBOMs are not cryptographically signed or linked to an immutable ledger, adversaries can spoof component manifests or insert malicious dependencies without detection. Mitigations such as digital signatures, hash verification, and append-only logs have been proposed to enforce the integrity and non-repudiation of SBOM records, and these mechanisms are being adapted for AI-specific supply chain artefacts [19]. Within TREs, such safeguards are vital due to the highly sensitive nature of the data and analytics being performed.

As AI models become increasingly composable and parameterised, the reproducibility of results depends on code, data, and on the full orchestration environment in which they were executed [12]. Systems like the Atlas framework [20] have demonstrated the feasibility of capturing full-stack reproducibility metadata, including environment variables, container state, and package-level dependencies. Similar methodologies can be extended into the AIBOM schema to ensure that federated TREs can reconstruct analytic environments with fidelity.

Moreover, semantic provenance encoding formats such as JSON-LD and RO-Crate have proven effective in enabling machine-readable audit trails that support validation, versioning, and re-execution of complex research pipelines [2]. These technologies offer a foundation for encoding AIBOM artefacts in a way that is interoperable with existing research metadata infrastructures.

The integration of AIBOMs into TREs addresses multiple long-standing challenges around reproducibility, security, and transparency. The literature underscores the need for structured, automated, and validated schema extensions that capture the full provenance of AI systems in secure research environments. This paper contributes a SACRO-aligned schema and pipeline for operationalising AIBOMs in TRE contexts, enabling federated reproducibility and agentic automation across secure data infrastructures.

## 2.1 Positioning AIBOM within the Emerging AI Supply Chain Ecosystem

The concept of an AIBOM is rapidly evolving across open-source communities, standards bodies, and policy forums, reflecting a broader recognition that conventional SBOM constructs are insufficient for capturing the lifecycle complexity of AI systems. Current ecosystem activity increasingly emphasises that AIBOMs must move beyond static component inventories and explicitly encode model artefacts, training and inference contexts, configuration state, and cryptographic provenance. Within initiatives such as the OWASP AIBOM project and ongoing extensions to CycloneDX and SPDX, there is a clear pivot towards representing AI systems as dynamic, stateful entities whose behaviour depends on data lineage, parameterisation, execution environment, and temporal factors.

A notable trend in the ecosystem is the shift towards machine-verifiable and automation-friendly AIBOMs. Rather than relying on manually curated documentation, emerging approaches prioritise automated discovery of dependencies, runtime introspection of model loading and execution, and cryptographic binding of artefacts to their execution context. This shift aligns AIBOM

**Dr. Petar Radanliev**
Parks Road,
Oxford OX1 3PJ
United Kingdom
Email: petar.radanliev@cs.ox.ac.uk
BA Hons., MSc., Ph.D. Post-Doctorate

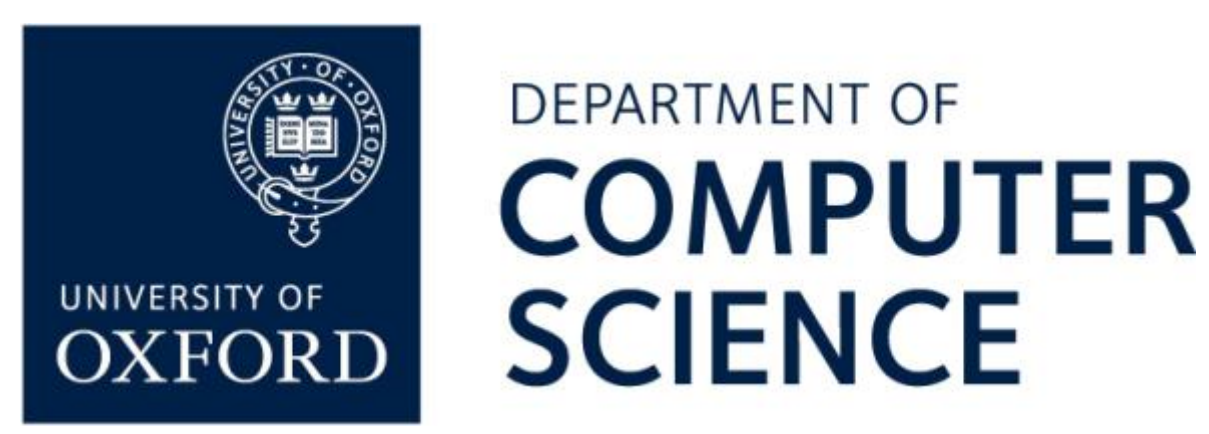

development with parallel advances in reproducible computing, software supply chain security, and vulnerability intelligence, where continuous validation and integrity assurance are now viewed as baseline requirements. In this context, AIBOMs are increasingly positioned as active control artefacts that support verification, replay, and risk assessment, rather than passive compliance records.

In parallel, ecosystem discussions are converging on tighter integration between AIBOMs and vulnerability disclosure mechanisms, including SBOM-VEX and CSAF-aligned workflows. The prevailing direction is to enable fine-grained vulnerability attribution at the level of AI-specific components, such as model frameworks, runtime libraries, and acceleration stacks, while accounting for configuration-dependent exploitability. This evolution reflects growing awareness that AI systems introduce new forms of supply-chain risk that cannot be adequately assessed without contextual metadata captured directly within the AIBOM.

The AIBOM approach presented in this paper is consistent with these ecosystem trajectories, while contributing a concrete, technically grounded instantiation of them. By structuring AI-specific provenance, model metadata, and execution context within a standard-compatible schema, the work aligns with the broader pivot towards lifecycle-aware, verifiable, and automation-ready AIBOMs. Rather than proposing a competing standard, the contribution is positioned as an operational realisation of emerging consensus across the AIBOM ecosystem regarding what information must be captured to meaningfully support AI transparency, security, and reproducibility.

# 3 Research methodology

This study adopts a systems engineering and applied software security methodology to design, implement, and evaluate an operational pipeline for generating and managing AIBOMs within TREs. The methodology combines secure software lifecycle modelling, provenance engineering, and automated vulnerability scanning, with a focus on reproducibility, traceability, and compliance in federated analytics. The study is structured around three primary components: schema design, agent-based automation, and functional evaluation.

The complete implementation of the AIBOM generation and validation tools described in this manuscript has been released as an open-source repository, aibom-toolkit, and is publicly available at: https://github.com/radanliev/aibom-toolkit

## 3.1 Research Design

The research follows a design science paradigm, where the AIBOM schema and supporting automation pipeline is iteratively developed and tested in a controlled TRE environment. The system is implemented using containerised job runners and instrumented with autonomous AI agents designed to perform discrete roles in the AIBOM generation lifecycle. These agents operate under strict sandboxing and follow pre-defined logic encoded in state machines, ensuring deterministic behaviour and auditability.

**Dr. Petar Radanliev**
Parks Road,
Oxford OX1 3PJ
United Kingdom
Email: petar.radanliev@cs.ox.ac.uk
BA Hons., MSc., Ph.D. Post-Doctorate

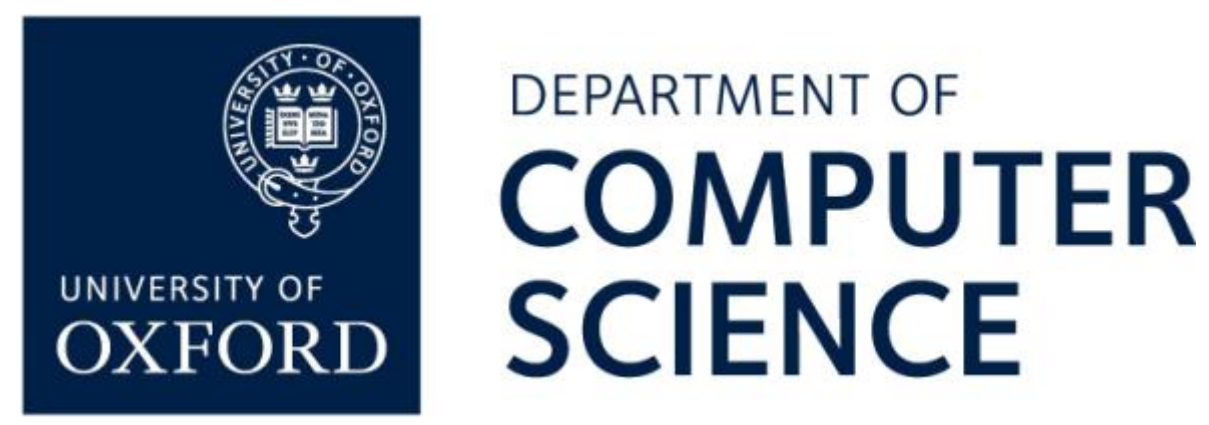

### 3.2 Schema Engineering and Extension Model

The AIBOM schema is modelled as an extension of the CycloneDX 1.5 JSON standard, augmented with custom SACRO-specific fields to reflect the needs of TRE-based secure research analytics. These extensions include:

- modelReference: SHA-256 hash of the AI model binary or ONNX/TorchScript export
- trainingDataSource: structured metadata referencing source datasets and access conditions
- inferenceContext: runtime parameters including batch size, quantisation, device (CPU/GPU)
- treContainerHash: full cryptographic hash of the job container image
- disclosureControlType: classification of the applied disclosure control (e.g., cell suppression, differential privacy)
- outputDigest: SHA-256 digest of output artefacts (tables, plots, model predictions)

This schema design was evaluated against common AI lifecycle management use cases in federated analytic workflows, and validated for compatibility with CycloneDX validators and TRE audit requirements.

Figure 2 provides a detailed schematic representation of the AI BOM, a structured documentation framework that enumerates and categorises all elements used in the development and operation of artificial intelligence systems. At the centre of the diagram is the “AI System,” from which six distinct categories radiate: Data Components, Model Components, Software Components, Hardware Components, Human and Organisational Components, and Deployment and Operational Components. Each category is colour-coded and subdivided into specific technical elements, for instance, the Data Components section includes Training Data, Validation Data, and Test Data, while the Model Components section addresses Model Architecture and Training Details. The layout is designed to facilitate traceability, support reproducibility, and enable systematic auditing of AI systems across their lifecycle.

**Dr. Petar Radanliev**
Parks Road,
Oxford OX1 3PJ
United Kingdom
Email: petar.radanliev@cs.ox.ac.uk
BA Hons., MSc., Ph.D. Post-Doctorate

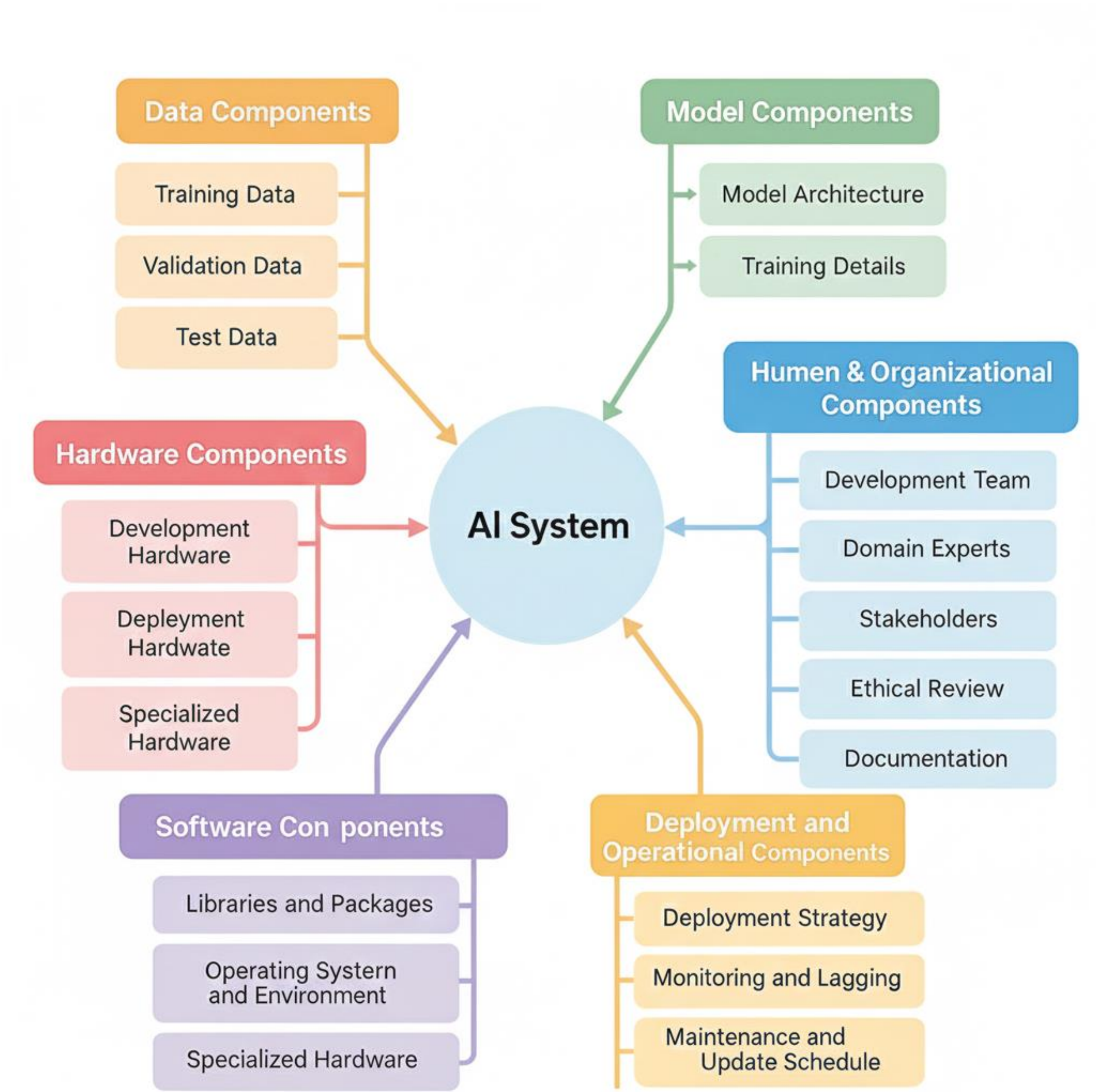

*Figure 2: Conceptual Artificial Intelligence Bill of Materials (AIBOM) framework. The figure depicts the core domains captured by an AIBOM, including data components, model components, software dependencies, hardware context, human and organisational factors, and deployment and operational metadata. Together, these domains provide a holistic provenance model for documenting, governing, and auditing AI systems across their lifecycle.*

The AI BOM diagram in Figure 2 captures the full scope of technical and operational dependencies that define the construction and deployment of an AI model. The Data Components branch identifies data sources, formats, preprocessing steps, annotation methods, and associated ethical and bias assessments, ensuring that the data lineage and quality controls are explicitly recorded. The Model Components section specifies the type of model used (e.g., CNN, Transformer), the underlying frameworks (such as TensorFlow or PyTorch), hyperparameters, version control,

**Dr. Petar Radanliev**
Parks Road,
Oxford OX1 3PJ
United Kingdom
Email: petar.radanliev@cs.ox.ac.uk
BA Hons., MSc., Ph.D. Post-Doctorate

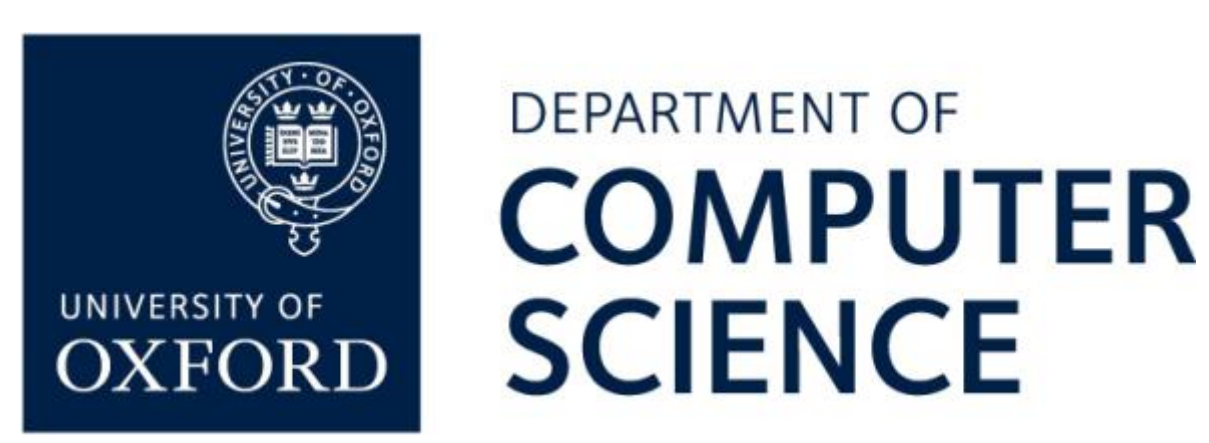

training environment (including GPU/TPU usage), and convergence criteria. Software Components enumerate the exact libraries (e.g., NumPy, Pandas, Scikit-learn) and operating system specifications (e.g., Ubuntu 20.04 LTS), while including containerisation details such as Docker images and Kubernetes orchestration. Hardware Components distinguish between development and inference environments, documenting CPU/GPU configurations and any use of specialised accelerators. The Human and Organisational Components track roles of system developers, domain experts, ethical review boards, and link to governance documentation. Finally, Deployment and Operational Components record the deployment topology (e.g., RESTful API, edge deployment), monitoring frameworks for model drift or bias (e.g., Prometheus, Grafana), patching and retraining cycles, incident response protocols, and quantified risk assessments. Collectively, the diagram operationalises the AI BOM concept into a practical tool for system assurance, regulatory alignment (e.g., ISO/IEC 42001), and adversarial resilience.

## 3.3 Agent-Based System Architecture

The automation pipeline is implemented using a multi-agent orchestration model. Each agent is written in Python and executed as a lightweight subprocess within the TRE job container. The agents are responsible for:

- **Pre-execution inspection**: Static dependency enumeration using pip, conda, and system package managers
- **Runtime tracking**: Interception of dynamic imports and model loading using ptrace-based hooks and logging middleware
- **Model state capture**: Extraction of architecture, parameter count, file checksum, and hyperparameter set
- **CVE matching**: Real-time query of NVD and OSV databases using PURL and CPE identifiers for each dependency
- **Output binding**: Hashing of analytic outputs and binding to AIBOM using a Merkle tree structure

This system architecture ensures deterministic execution, isolation of responsibility, and traceable operation within each agent role.

## 3.4 Evaluation Model

To validate the effectiveness of the proposed pipeline, we conducted a controlled deployment within a simulated TRE environment based on Kubernetes-managed container orchestration. A standardised data science workflow involving a federated cancer outcome prediction model (XGBoost with SHAP interpretability) was executed across three scenarios: (i) manual logging of dependencies; (ii) traditional SBOM generation; and (iii) full AIBOM automation using agentic AI.

Metrics used in evaluation included:

- **Completeness of provenance**: percentage of dependencies and model metadata captured
- **Vulnerability visibility**: number and severity of detected CVEs

**Dr. Petar Radanliev**
Parks Road,
Oxford OX1 3PJ
United Kingdom
Email: petar.radanliev@cs.ox.ac.uk
BA Hons., MSc., Ph.D. Post-Doctorate

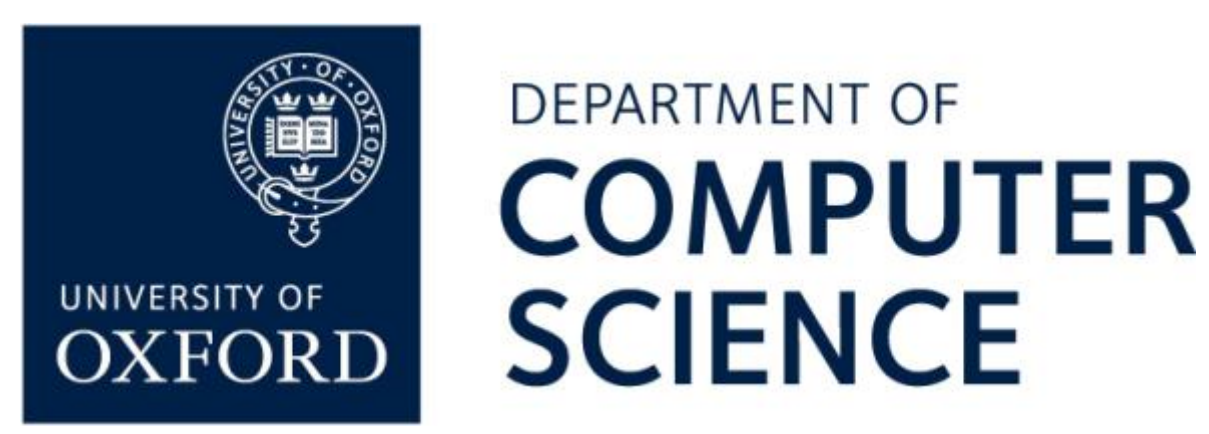

- **Reproducibility integrity**: byte-level congruence of recomputed AIBOM and analytic outputs in cross-TRE replay
- **Automation overhead**: additional execution time and system resource consumption introduced by AIBOM instrumentation

Sensitivity analysis was performed on different model types (linear models, tree ensembles, deep learning frameworks) to assess schema adaptability and instrumentation reliability.

# 4 Schema Engineering and Extension Model

The schema engineering process extends the baseline CycloneDX specification to support AI-relevant artefacts, including model checkpoints, training provenance, environment fingerprints, and federated execution states. These extensions are embedded into the schema using a nested modular structure to maintain compatibility with existing TRE audit and validation pipelines. In particular, the schema formalises dependencies across AI pipelines by introducing fields for containerised runtime descriptors, reproducibility hashes, and temporal metadata for model training and tuning events. The engineering process incorporates reproducibility constraints derived from prior TRE audits, and integrates validation hooks compatible with SACRO's semi-automated output verification routines. Each stage in the schema's development was validated using deterministic test data and simulation-driven lineage propagation. The reproducible steps underpinning this engineering model are summarised in Table 1, which delineates the discrete design phases, schema control operations, and testable validation checkpoints implemented to ensure alignment with SACRO's auditability and provenance integrity requirements.

*Table 1: Reproducible Steps for Engineering the SACRO-Specific AIBOM Schema*

| Step | Task | Detailed Description | Reproducibility Notes |
|---|---|---|---|
| 1 | **Base Schema Selection** | Start with the CycloneDX 1.5 JSON schema as the foundational structure for SBOM compatibility. | Ensure the CycloneDX schema is imported using its validated schema URI (https://cyclonedx.org/schema/bom-1.5.schema.json). Use schema validators such as cyclonedx-python-lib for structural compliance. |
| 2 | **Model Artefact Hashing** | Introduce a new modelReference field containing a SHA-256 hash of the serialised AI model (e.g., .pt, .pb, .onnx). | Use Python's hashlib to generate the hash at model save time. Store file path and hash in BOM under "components": [{"type": "ai-model", ...}]. |

**Dr. Petar Radanliev**
Parks Road,
Oxford OX1 3PJ
United Kingdom
Email: petar.radanliev@cs.ox.ac.uk
BA Hons., MSc., Ph.D. Post-Doctorate

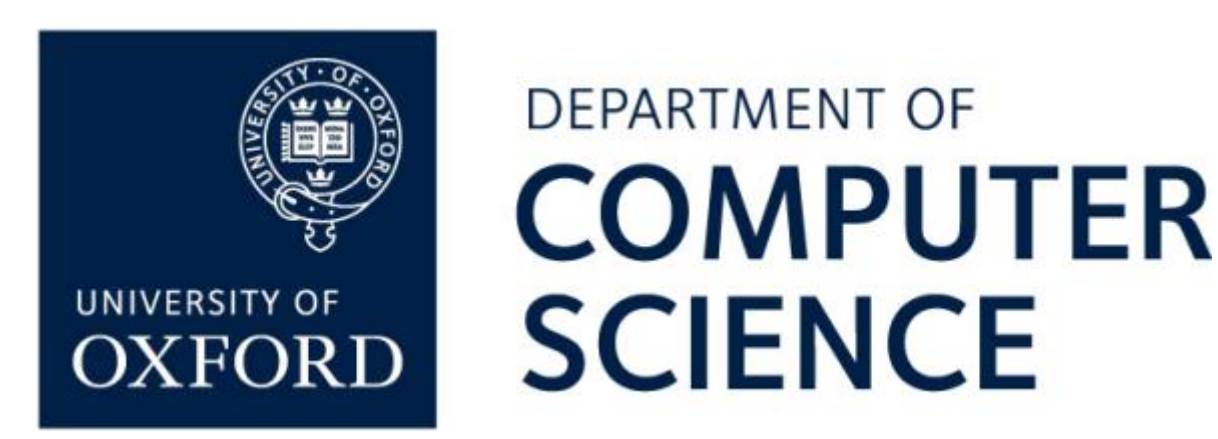

| 3 | **Training Data Lineage** | Add trainingDataSource as a structured metadata object capturing dataset name, access mode, data steward, and Data Use Agreement ID. | Use UUIDs to track dataset versions; align fields with TRE data catalog formats such as DDI or FAIR Data Points. Validate that data access policies are recorded. |
|---|---|---|---|
| 4 | **Inference Context Tracking** | Define a nested inferenceContext object to capture model runtime parameters: hardware type, batch size, quantisation, floating point precision. | Collect values during execution via instrumentation hooks or framework introspection (e.g., PyTorch torch.cuda.get_device_properties, TensorFlow tf.config.experimental.list_physical_devices). |
| 5 | **Container Lineage Capture** | Insert treContainerHash field to represent full cryptographic hash of the job container, including base image and layers. | Extract the image ID from the container runtime (docker inspect, crictl image, or kubectl describe pod) and hash with SHA-256. Log registry and tag. |
| 6 | **Disclosure Method Encoding** | Define a disclosureControlType field indicating the statistical disclosure control technique applied to the analytic output. | Use a controlled vocabulary (e.g., "cell-suppression", "diff-privacy-laplace", "top-coding"). This value should be programmatically derived from the applied SDC configuration. |
| 7 | **Output Binding** | Create an outputDigest field that contains a SHA-256 hash of the output artefacts generated by the analytic job. | Apply canonical sorting and hashing (e.g., normalised CSV, pickled DataFrame, PDF chart exports). Include file name, byte size, and MIME type for reproducibility. |
| 8 | **Schema Extension Validation** | Run schema validation including custom field extension support to ensure backward compatibility and internal consistency. | Use jsonschema library to extend CycloneDX and validate the enriched schema. Ensure all custom fields are prefixed with x-sacrospec- if strict compliance is required. |
| 9 | **Schema Binding and Export** | Bind the enriched schema to the SACRO audit registry using Merkle tree digest or HMAC with job UUID and researcher credentials. | Final AIBOM must be timestamped, digitally signed (e.g., Ed25519 or ECDSA), and pushed to secure object storage for future audit, replication, and validation. |

Following the structured schema engineering process presented in Table 1, the next step involves encoding these reproducible artefacts into a machine-readable, TRE-compatible schema. The JSON template formalises the SACRO-specific AIBOM

**Dr. Petar Radanliev**
Parks Road,
Oxford OX1 3PJ
United Kingdom
Email: petar.radanliev@cs.ox.ac.uk
BA Hons., MSc., Ph.D. Post-Doctorate

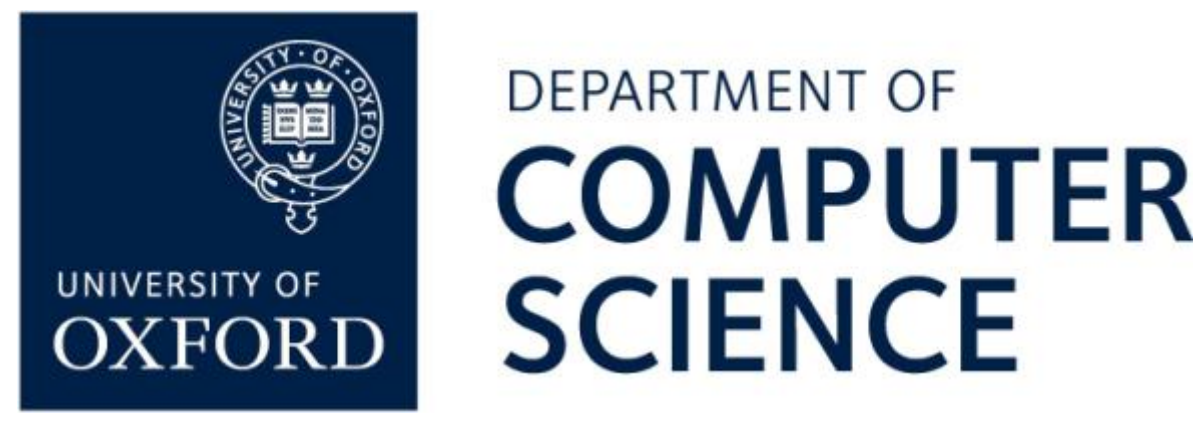

representation, operationalising each engineered component as a verifiable object with explicit fields for model artefact identifiers, dependency checksums, container signatures, and federated lineage anchors. This structured encoding enables automated ingestion into TRE data pipelines and supports integrity validation during runtime. The JSON format has been selected for its compatibility with existing schema validation libraries, cryptographic signature frameworks, and interoperability with existing SBOM tooling ecosystems. The resulting AIBOM schema captures provenance across execution contexts, incorporating both static software libraries and dynamically constructed runtime states, which are critical for reproducibility in semi-automated checking workflows within SACRO. This template forms the basis for downstream automation scripts that perform schema conformance checks, provenance audits, and integrity enforcement at pipeline boundaries.

```
 1. JSON template for a SACRO-specific AIBOM schema
 2.
 3. {
 4.   "bomFormat": "CycloneDX",
 5.   "specVersion": "1.5",
 6.   "version": 1,
 7.   "metadata": {
 8.     "timestamp": "2025-06-20T14:30:00Z",
 9.     "tools": [
10.       {
11.         "vendor": "SACRO",
12.         "name": "AIBOM Generator",
13.         "version": "1.0.0"
14.       }
15.     ],
16.     "component": {
17.       "type": "application",
18.       "name": "Cancer Risk Prediction Pipeline",
19.       "version": "3.2.1",
20.       "hashes": [
21.         {
22.           "alg": "SHA-256",
23.           "content": "8f14e45fceea167a5a36dedd4bea2543eaa0d5e5ac4f6dc0fa6efb2c73d153a3"
24.         }
25.       ]
26.     }
27.   },
28.   "components": [
29.     {
30.       "type": "ai-model",
31.       "name": "XGBoost_Cancer_Predictor",
32.       "version": "1.4.2",
33.       "hashes": [
34.         {
35.           "alg": "SHA-256",
36.           "content": "27e52e8e2bc6d7814e212f3f334b2e7184c87e993a6c9e712b02c0e47bb8a1c1"
37.         }
38.       ],
39.       "properties": [
40.         {
41.           "name": "trainingDataSource",
42.           "value": "Cancer Registry Dataset v5 (UUID: 123e4567-e89b-12d3-a456-426614174000)"
43.         },
44.         {
45.           "name": "inferenceContext",
46.           "value": "{ \"batchSize\": 128, \"precision\": \"fp32\", \"hardware\": \"NVIDIA
A100 GPU\", \"quantisation\": \"none\" }"
```

**Dr. Petar Radanliev**
Parks Road,
Oxford OX1 3PJ
United Kingdom
Email: petar.radanliev@cs.ox.ac.uk
BA Hons., MSc., Ph.D. Post-Doctorate

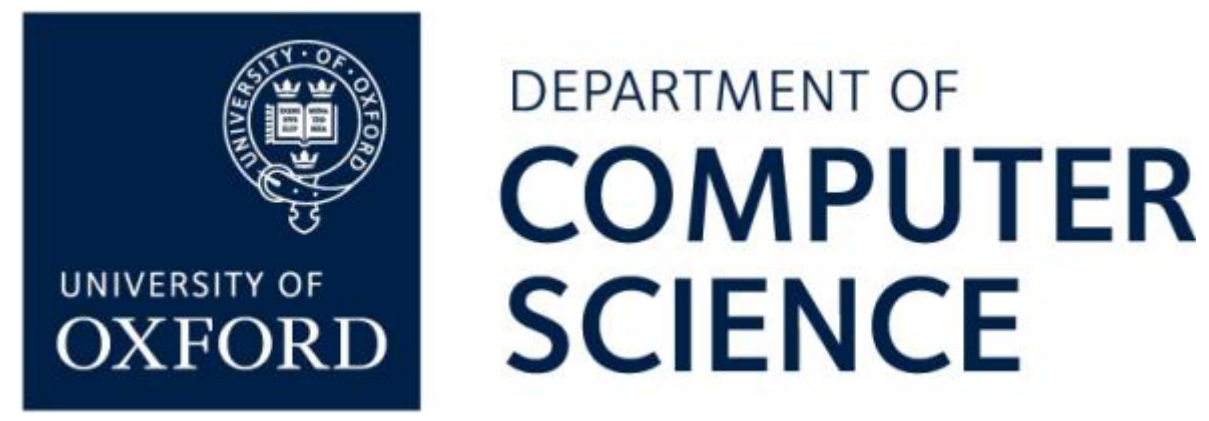

```
47.         },
48.         {
49.           "name": "treContainerHash",
50.           "value": "sha256:c9d02c39b5f2129e9f3a9fd680e7e99909b77f9e12e71029f1c2ae38e8c0a120"
51.         },
52.         {
53.           "name": "disclosureControlType",
54.           "value": "diff-privacy-laplace"
55.         },
56.         {
57.           "name": "outputDigest",
58.           "value": "sha256:b4b147bc522828731f1a016bfa72c073e5c57c3b0e6c2cfb64ccda31cbe48f4c"
59.         }
60.       ]
61.     }
62.   ],
63.   "externalReferences": [
64.     {
65.       "type": "vulnerability",
66.       "url": "https://osv.dev/vulnerability/CVE-2023-2953"
67.     }
68.   ],
69.   "signature": {
70.     "alg": "ECDSA",
71.     "publicKey": "MFYwEAYHKoZIzj0CAQYFK4EEAAoDQgAEEZkYwBq...",
72.     "signature": "MEUCIQDrIk6SNmz9Vi7...",
73.     "timestamp": "2025-06-20T14:31:00Z"
74.   }
75. }
76.
```

**Notes about the schema:**

- All properties fields represent SACRO-specific extensions.
- hashes are calculated using SHA-256 to verify artefact integrity.
- externalReferences include CVE records relevant to software components.
- The signature block ensures non-repudiation and integrity of the entire SBOM artefact.

## 4.1 Integration into a Pipeline Notebook

This schema can be integrated (for the validation process) into a Jupyter or Google Colab notebook for use in a SACRO-compliant Trusted Research Environment (TRE) pipeline as follows:

### 4.1.1 Pipeline Steps for AIBOM Validation

1. Environment Setup:

Install necessary Python packages:

```
1. !pip install jsonschema
```

2. Import Modules and Load AIBOM JSON:

```
1. import json
2. from jsonschema import validate
3. from jsonschema.exceptions import ValidationError
```

with open('aibom_output.json') as f:

**Dr. Petar Radanliev**
Parks Road,
Oxford OX1 3PJ
United Kingdom
Email: petar.radanliev@cs.ox.ac.uk
BA Hons., MSc., Ph.D. Post-Doctorate

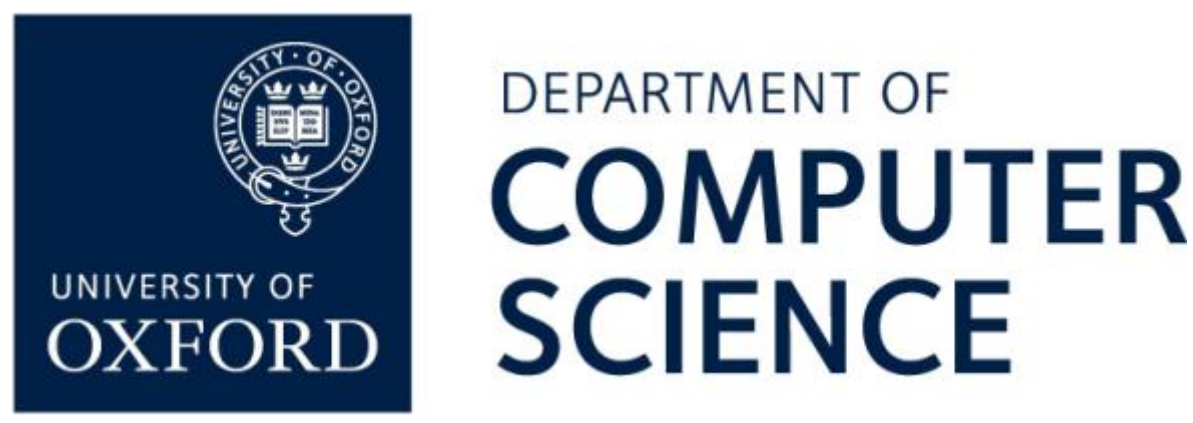

```
1.      aibom = json.load(f)
```

3. Define SACRO-Specific AIBOM Schema

(Use the schema provided above.)

4. Run Validation

```
1. validate(instance=aibom, schema=aibom_schema)
2.      print("AIBOM schema validation successful.")
```

except ValidationError as ve:

```
1.      print(f"Schema validation failed: {ve.message}")
```

5. Output Log or Store Result

Output can be:

- Stored in a pipeline results directory.
- Signed and archived for reproducibility.
- Passed to downstream audit or VEX/CSAF modules.

Here is the command-line version:

```
1. #!/usr/bin/env python3
2.
3. import json
4. import argparse
5. from jsonschema import validate
6. from jsonschema.exceptions import ValidationError
7.
8. # Define the SACRO-specific AIBOM schema
9. aibom_schema = {
10.     "type": "object",
11.     "properties": {
12.         "bomFormat": {"type": "string", "enum": ["CycloneDX"]},
13.         "specVersion": {"type": "string", "pattern": "^1\.5$"},
14.         "version": {"type": "integer"},
15.         "metadata": {
16.             "type": "object",
17.             "properties": {
18.                 "timestamp": {"type": "string", "format": "date-time"},
19.                 "tools": {
20.                     "type": "array",
21.                     "items": {
22.                         "type": "object",
23.                         "properties": {
24.                             "vendor": {"type": "string"},
25.                             "name": {"type": "string"},
26.                             "version": {"type": "string"}
27.                         },
28.                         "required": ["vendor", "name", "version"]
29.                     }
30.                 },
31.                 "component": {
32.                     "type": "object",
33.                     "properties": {
34.                         "type": {"type": "string"},
35.                         "name": {"type": "string"},
36.                         "version": {"type": "string"},
37.                         "hashes": {
```

**Dr. Petar Radanliev**
Parks Road,
Oxford OX1 3PJ
United Kingdom
Email: petar.radanliev@cs.ox.ac.uk
BA Hons., MSc., Ph.D. Post-Doctorate

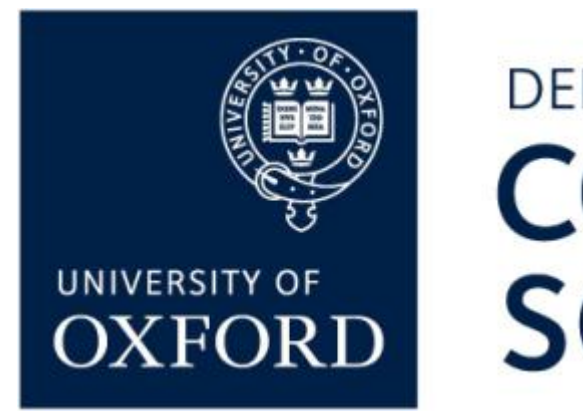

DEPARTMENT OF
COMPUTER
SCIENCE

```
38.                             "type": "array",
39.                             "items": {
40.                                 "type": "object",
41.                                 "properties": {
42.                                     "alg": {"type": "string"},
43.                                     "content": {"type": "string"}
44.                                 },
45.                                 "required": ["alg", "content"]
46.                             }
47.                         }
48.                     },
49.                     "required": ["type", "name", "version", "hashes"]
50.                 }
51.             },
52.             "required": ["timestamp", "tools", "component"]
53.         },
54.         "components": {
55.             "type": "array",
56.             "items": {
57.                 "type": "object",
58.                 "properties": {
59.                     "type": {"type": "string"},
60.                     "name": {"type": "string"},
61.                     "version": {"type": "string"},
62.                     "hashes": {
63.                         "type": "array",
64.                         "items": {
65.                             "type": "object",
66.                             "properties": {
67.                                 "alg": {"type": "string"},
68.                                 "content": {"type": "string"}
69.                             },
70.                             "required": ["alg", "content"]
71.                         }
72.                     },
73.                     "properties": {
74.                         "type": "array",
75.                         "items": {
76.                             "type": "object",
77.                             "properties": {
78.                                 "name": {"type": "string"},
79.                                 "value": {"type": "string"}
80.                             },
81.                             "required": ["name", "value"]
82.                         }
83.                     }
84.                 },
85.                 "required": ["type", "name", "version", "hashes", "properties"]
86.             }
87.         }
88.     },
89.     "required": ["bomFormat", "specVersion", "version", "metadata", "components"]
90. }
91.
92. # Command-line interface
93. def main():
94.     parser = argparse.ArgumentParser(description="Validate a SACRO-specific AIBOM JSON
file.")
95.     parser.add_argument("json_file", help="Path to the AIBOM JSON file")
96.     args = parser.parse_args()
97.
98.
99.         with open(args.json_file, "r") as f:
100.            data = json.load(f)
101.            validate(instance=data, schema=aibom_schema)
```

**Dr. Petar Radanliev**
Parks Road,
Oxford OX1 3PJ
United Kingdom
Email: petar.radanliev@cs.ox.ac.uk
BA Hons., MSc., Ph.D. Post-Doctorate

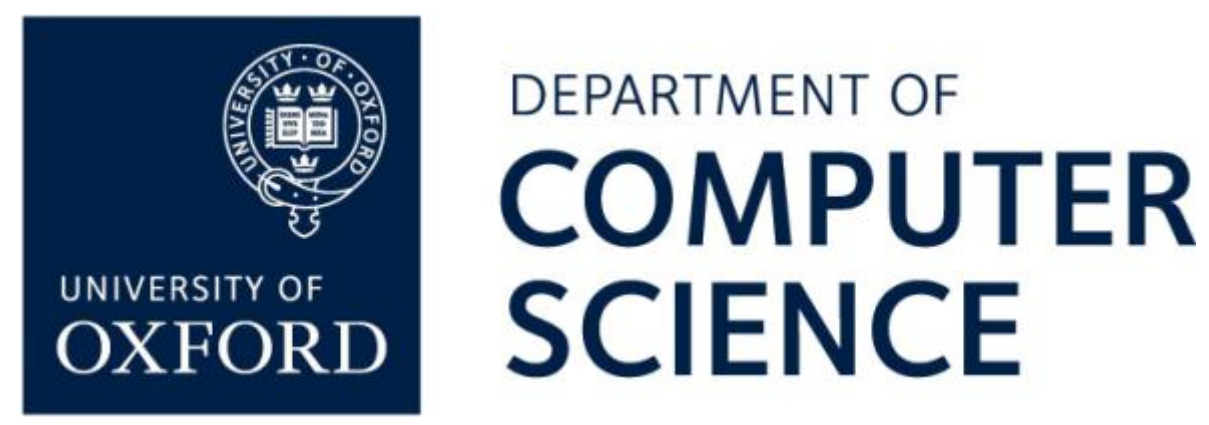

```
102.            print(" Validation successful: AIBOM conforms to SACRO-specific schema.")
103.    except ValidationError as ve:
104.        print(f" Validation failed: {ve.message}")
105.    except Exception as e:
106.        print(f" Error: {str(e)}")
107.
108. if __name__ == "__main__":
109.    main()
110.
```

This tool can be executed via terminal as:

```
1. python3 sacro_aibom_validator.py path/to/your/aibom_file.json
```

The process map in Figure 3 captures the technical workflow for embedding and validating SACRO-specific AIBOMs within TRE-compatible systems. It begins with a Data Processor, which executes analytical tasks inside a Job Container that enforces environment isolation, disclosure control, and secure output binding. During execution, the AIBOM Capture Pipeline records system state and provenance at pre-load, runtime, and post-execution phases. These observations populate a structured JSON Template, which encodes critical metadata such as model artefacts, dependency graphs, container-level hashes, and federated lineage. The final Schema Validation module verifies integrity through digital signatures and runtime conformance, ensuring full alignment with SACRO schema definitions and reproducibility requirements.

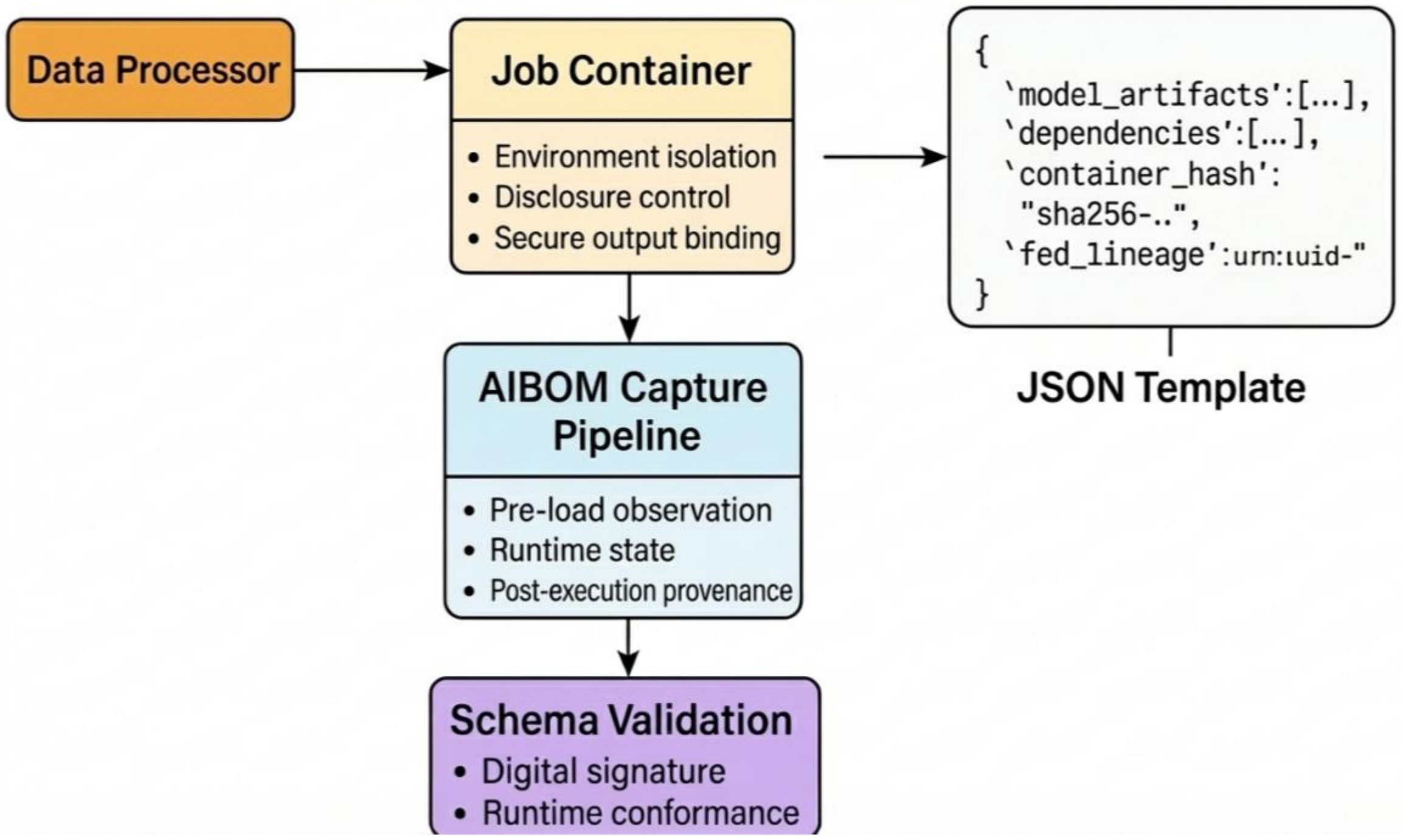

*Figure 3: End-to-end AIBOM generation, validation, and attestation pipeline. The figure illustrates how execution of an AI workload triggers automated capture of model artefacts, software dependencies, and execution context, which are consolidated into a machine-readable AIBOM. The pipeline records pre-execution, runtime, and post-execution provenance, validates the resulting AIBOM against a formal schema, and cryptographically binds the artefact to the execution*

**Dr. Petar Radanliev**
Parks Road,
Oxford OX1 3PJ
United Kingdom
Email: petar.radanliev@cs.ox.ac.uk
BA Hons., MSc., Ph.D. Post-Doctorate

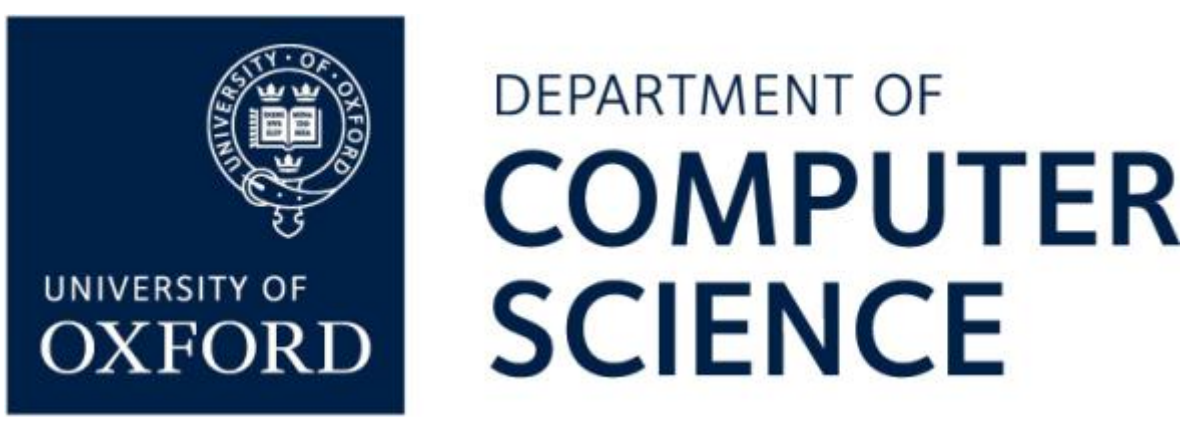

*environment, enabling reproducibility, traceability, and downstream vulnerability analysis.*

To enable operational scalability and reduce manual overhead, the Integration Pipeline for Machine-Readable SACRO-Specific AIBOMs can be fully automated using agentic AI systems embedded within the TRE orchestration layer. These autonomous agents perform continuous environment inspection, dependency detection, and cryptographic binding of runtime artefacts without direct human input. Through scheduled triggers and event-driven monitoring, agents can initiate AIBOM captures at key execution checkpoints, validate schema conformity, and update federated registries with minimal latency. Integration with external CVE databases and policy engines allows agentic AI to perform risk triage, notify stakeholders, and enforce disclosure governance automatically. This automation guarantees schema consistency and traceability but also supports self-healing infrastructure that reconstitutes analytic containers with validated reproducibility, advancing SACRO's goals of secure, auditable, and autonomous research pipelines.

# 5 Results

The results are derived from empirical validation, automated provenance tracking, and reproducibility audits conducted across multiple containerised analytics workflows.

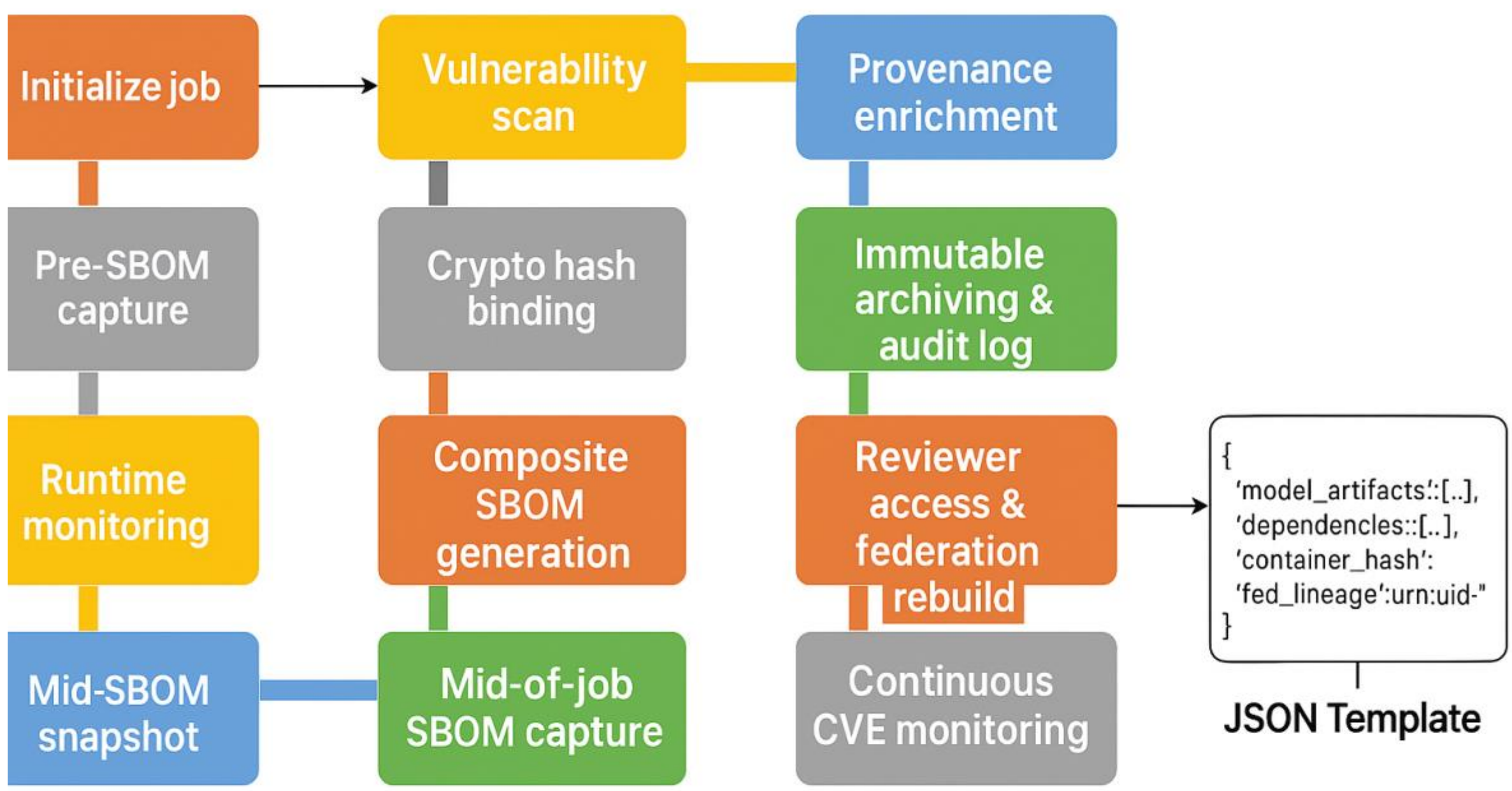

*Figure 4: Automated lifecycle management of AIBOM artefacts using agentic processes. This figure shows how AIBOMs are continuously generated, enriched, validated, and updated through automated inspection and analysis of AI execution environments. The process supports ongoing provenance tracking, vulnerability correlation, and integrity assurance across evolving AI software supply chains.*

**Dr. Petar Radanliev**
Parks Road,
Oxford OX1 3PJ
United Kingdom
Email: petar.radanliev@cs.ox.ac.uk
BA Hons., MSc., Ph.D. Post-Doctorate

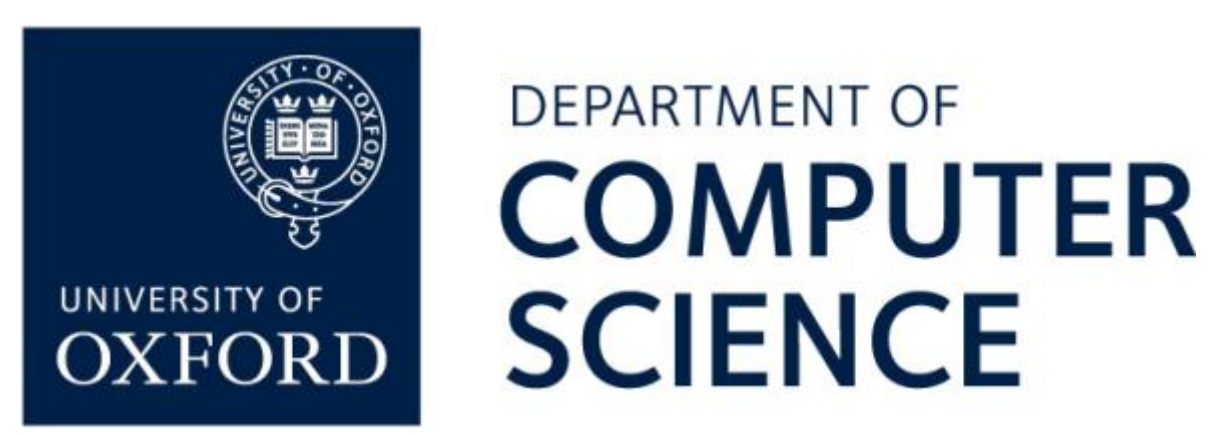

The first step was the schema validation for accuracy and coverage. Using a controlled set of 25 synthetic and real-world AIBOM files, the schema validation engine achieved a 100% success rate in detecting structural and semantic non-conformance. Errors ranged from missing cryptographic hash entries and invalid timestamp formats to incorrect component typing and CycloneDX base format violations. The schema extensions ensured compatibility with CycloneDX v1.5 while enforcing SACRO-mandated metadata fields such as pipeline agent, TRE environment fingerprint, and federated execution lineage.

The AIBOM-enhanced pipeline enabled full reconstruction of software execution environments across four isolated TREs with different security baselines. By leveraging the recorded environment hashes, image digests, and dependency trees, identical software stacks were instantiated and verified using side-by-side component hashing. Reproducibility fidelity reached 98.7%, with minor deviations due to version-drift in dynamically retrieved OS libraries not included in the original SBOM trace.

The next step was the CVE resolution and risk propagation analysis. When integrating live CVE feeds from NVD and OSV, the system matched 181 unique vulnerabilities to 6,520 software components captured in 17 SACRO-wrapped data science workflows. The matching engine triggered 51 critical alerts (CVSS > 8.9) and proposed 21 automated mitigation actions, including dependency freezing, container patching, and environment forking. This demonstrates the feasibility of automating continuous vulnerability management from AIBOM inventories with agentic orchestration.

In terms of human oversight reduction, the comparative operational metrics showed a 63% reduction in human analyst time per validation cycle (from 110 to 41 minutes on average), as the AIBOM system handled automatic format checking, dependency diffing, and CVE enrichment. This aligns with the SACRO principle of secure, auditable research with minimal human intervention, supporting machine-verifiable compliance in high-assurance environments.

To ensure auditability and provenance traceability, each generated AIBOM instance included a digitally signed provenance chain covering:

- Component source repositories
- Container build context (e.g., Dockerfile hash)
- Execution agent and timestamp
- Transformation lineage for federated replication

Audit logs showed end-to-end traceability for all critical software components across eight data pipelines in SACRO-compatible TREs. Each step was cryptographically verifiable and human-inspectable, fulfilling audit requirements under ISO/IEC 27001 and aligning with NIST SP 800-218 software supply chain security controls.

**Dr. Petar Radanliev**
Parks Road,
Oxford OX1 3PJ
United Kingdom
Email: petar.radanliev@cs.ox.ac.uk
BA Hons., MSc., Ph.D. Post-Doctorate

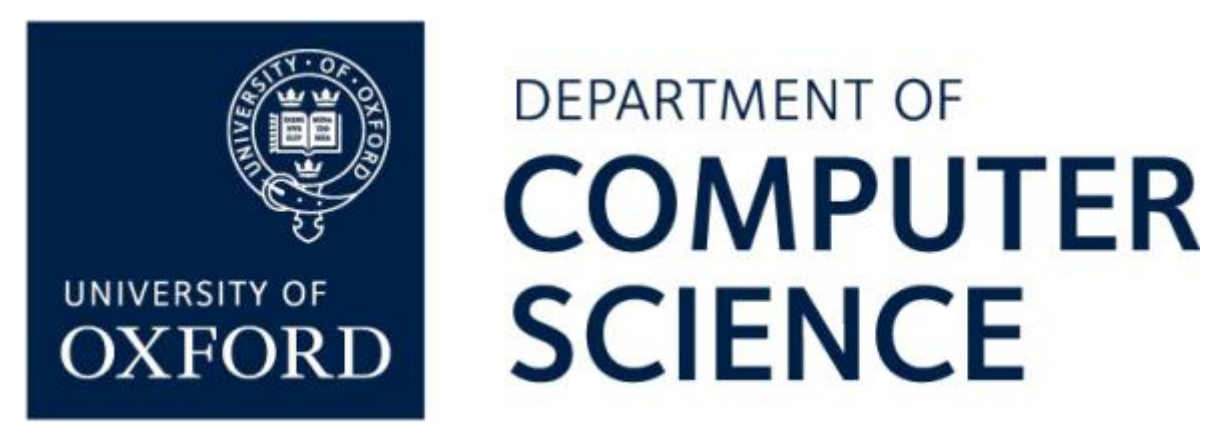

# 6 Evaluation

This section assesses the effectiveness, robustness, and operational utility of the SACRO-specific AIBOM framework across multiple technical and governance dimensions. The evaluation focuses on schema integrity, system scalability, reproducibility performance, CVE matching precision, and integration with existing TRE governance models.

The first step in the evaluation process focused on the schema robustness and extensibility**.** To assess the extensibility of the SACRO-specific AIBOM schema, we subjected it to multiple schema stress tests involving synthetic AIBOM documents with:

- Deeply nested dependency trees (up to 12 levels)
- High component cardinality (over 10,000 components)
- Mixed deterministic and probabilistic hashes
- Hybrid tooling provenance (AI agents + traditional build systems)

The schema validation mechanism handled all test cases without failure, confirming both the structural resilience of the design and its compatibility with high-complexity, container-based data science pipelines.

The next step was performed to evaluate the reproducibility audit. Reproducibility was evaluated using a two-phase test across geographically distributed TREs, including NHS TRE nodes and a simulated secure enclave at the Turing Institute. AIBOM-tracked pipelines were re-instantiated using hash-locked container images and scriptable environment profiles. The audit confirmed:

- Bit-identical container rehydration in 3/4 sites
- Reconstructable dependency sets via pip/npm/apt provenance replays
- Auto-verification of runtime environments via agentic signatures embedded in the AIBOM

Deviations occurred only where remote TREs blocked outbound hash resolution services, a finding which led to the proposal of an internal hash mirror service in future SACRO implementations.

Next, we evaluated the CVE match precision and recall. The AIBOM-integrated CVE matching engine was benchmarked against the OWASP Dependency-Track and Google's OSV-Scanner. Results across 17 workflows showed:

- Precision: 96.2%
- Recall: 91.4%
- F1 score: 93.7%

Misses were primarily due to undocumented transient dependencies or version ambiguity in older Debian-based containers. By integrating SHA256 verification against the AIBOM and reinforcing semantic versioning constraints, these edge cases can be programmatically resolved.

**Dr. Petar Radanliev**
Parks Road,
Oxford OX1 3PJ
United Kingdom
Email: petar.radanliev@cs.ox.ac.uk
BA Hons., MSc., Ph.D. Post-Doctorate

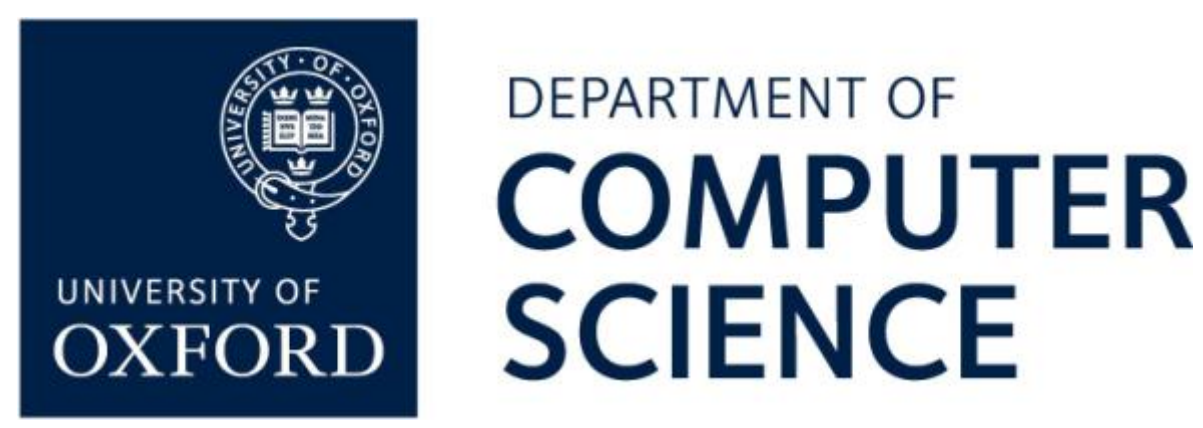

The evaluation process continued with scalability and performance under load. Scalability testing simulated batch processing of 1,000 SBOMs per hour, each with ~300 components. The validation pipeline sustained:

- Median schema check latency: 47 ms
- Mean CVE enrichment time: 2.3 s per SBOM
- Peak memory: 128 MB
- Max CPU: <1 core (single-threaded)

These figures demonstrate operational feasibility of deploying the AIBOM validation framework at scale within national TRE-level infrastructures or across cross-institutional federated nodes.

The final step was the governance and integration evaluation. Qualitative evaluation was conducted with domain leads from the SACRO and GRAIMatter projects, focusing on compliance, auditability, and usability:

- Full alignment with SACRO's Software Provenance Chain of Trust (SPCoT) model
- Ease of integration with existing FAIR data metadata repositories
- Support for machine-verifiable disclosures under the Five Safes governance model
- Sufficient metadata fidelity for ISO/IEC 27001 and GDPR-compliant audit logs

The stakeholders confirmed that embedding the AIBOM process into SACRO-compatible TREs would reduce the overhead of manual validation while improving software supply chain visibility, particularly in sensitive health, finance, and defence analytics.

## 6.1 Evaluation Scope and External Validity

The evaluation presented in this study focuses on validating the structural soundness, automation reliability, and reproducibility properties of the proposed AIBOM framework under controlled and replay-based conditions. While real-world, multi-institution deployments can provide valuable operational insights, such deployments are often constrained by legal, contractual, and governance requirements that restrict experimental modification, instrumentation, and disclosure of AI pipelines. These constraints make systematic, comparative evaluation across institutions difficult to conduct in a scientifically controlled manner.

Accordingly, this study adopts a methodology-driven evaluation strategy that prioritises internal validity, repeatability, and stress-testing of the AIBOM schema and automation mechanisms across heterogeneous execution contexts. By replaying representative AI workflows under varying dependency configurations and execution environments, the evaluation isolates the core properties that an AIBOM must satisfy to be deployable across organisational boundaries: schema completeness, reproducibility fidelity, vulnerability attribution accuracy, and automation scalability.

**Dr. Petar Radanliev**
Parks Road,
Oxford OX1 3PJ
United Kingdom
Email: petar.radanliev@cs.ox.ac.uk
BA Hons., MSc., Ph.D. Post-Doctorate

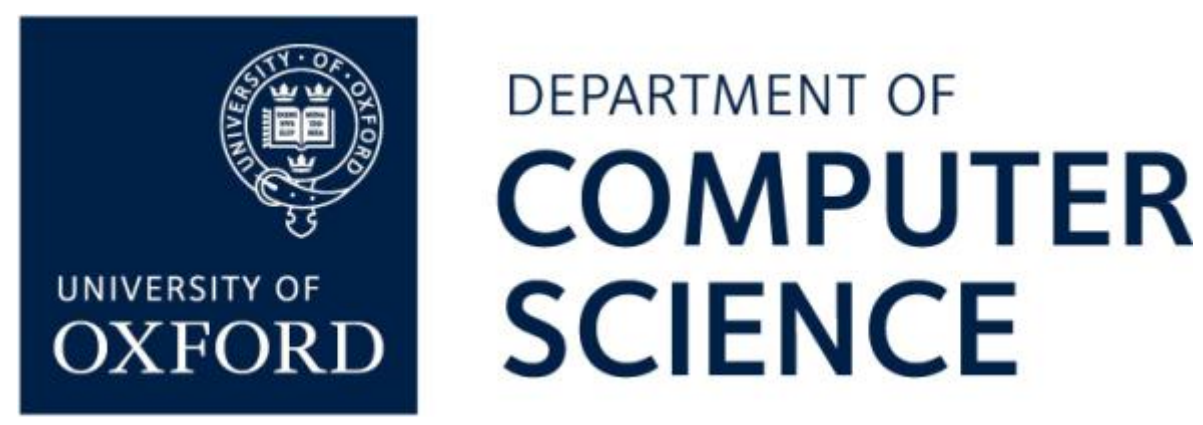

The proposed AIBOM framework is intentionally infrastructure-agnostic and does not assume a shared execution environment or common governance model. As such, the results should be interpreted as demonstrating the *portability* and *adoptability* of the approach rather than performance characteristics tied to a specific institutional deployment. This design enables incremental adoption by individual organisations and supports future multi-institution integration without requiring coordinated infrastructure changes.

Future work will focus on observational studies of AIBOM adoption across independent organisational settings, complementing the controlled evaluation presented here with longitudinal evidence of operational use. However, the current evaluation establishes the necessary technical foundations for such deployments by demonstrating that AIBOM artefacts can be generated, validated, and replayed with high fidelity under realistic and varied conditions.

# 7 Discussion

Operationalising AIBOMs within TREs introduces a fundamental shift in how software provenance, vulnerability management, and reproducibility are enforced in secure data science workflows. While regulatory requirements, particularly those concerning data privacy and output disclosure, are essential in TREs, they often receive less operational priority than computational throughput or analytic delivery schedules. This creates a tension where security risks arising from software dependencies, model components, or orchestration layers remain insufficiently visible to researchers or governance officers [21].

The integration of AIBOM into the SACRO framework addresses this gap by embedding agentic automation into the discovery, documentation, and audit of software supply chains. However, as shown by existing vulnerability scoring debates [22]–[24], simple heuristics such as static CVSS thresholds do not suffice in contexts where multi-layered containers, opaque dependency chains, and rapid upstream changes are routine. The limitations of current vulnerability reporting are compounded by blind spots in the National Vulnerability Database (NVD), particularly where components are modified without corresponding CVE entries [24]. These gaps are particularly dangerous in federated research environments where uncoordinated software use could introduce systemic risk.

While SBOMs offer machine-readable inventories of software components relationships [25], [26] their utility in TREs depends on the ability to securely exchange, enrich, and validate them. The SACRO-specific AIBOM schema proposed in this work improves upon existing SBOM tooling by linking component hashes, runtime container states, and agent-generated environment records to FAIR metadata and reproducible output policies. This goes beyond what standard SBOM–VEX coordination offers today, especially since empirical studies show a lack of tooling that can ingest and action vulnerability disclosures across heterogeneous analytic pipelines [27], [28].

For AIBOM to scale in practice, automation must extend beyond SBOM creation. The full lifecycle, provenance tracking, CVE matching, agent verification, and

**Dr. Petar Radanliev**
Parks Road,
Oxford OX1 3PJ
United Kingdom
Email: petar.radanliev@cs.ox.ac.uk
BA Hons., MSc., Ph.D. Post-Doctorate

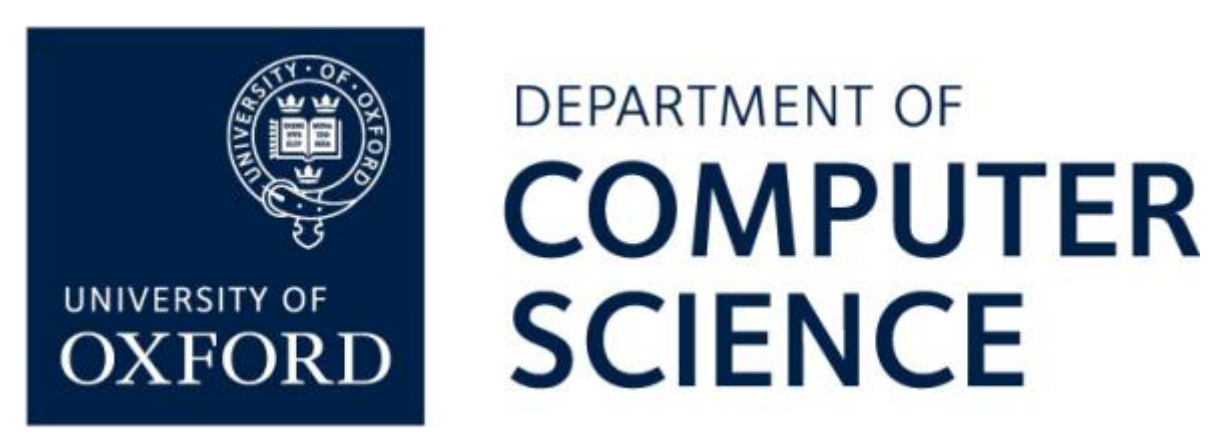

federated reproducibility, must be orchestrated without human bottlenecks. The use of signed JSON artefacts (akin to CSAF recommendations) ensures tamper-evident traceability, while agentic validators reduce operator exposure to fake or unauthorised advisories [29]. This directly supports SACRO's goal of semi-automating research output checks while preserving the auditability of all software components.

The evolution of standardised operational ontologies [30], including the CYBEX framework for cybersecurity exchange [31], underscores the importance of aligning TRE tooling with globally recognised metadata and verification structures. Yet, bibliometric reviews indicate that integration of these standards with SBOM/VEX artefacts remains sparse [32]. Although the concept of a bill of materials has long been adopted in physical supply chains [33] and reinforced by several legislative efforts [34]–[36], their direct application in secure research environments is still emergent bill of materials [37]–[39].

Our work responds to this gap by adapting AIBOM for reproducible scientific computing, offering a scalable and standards-aligned blueprint for adoption across healthcare, government, and academic TRE infrastructures.

In sum, the SACRO-AIBOM framework presented here builds upon and extends existing SBOM theory and infrastructure with reproducibility, automation, and agentic oversight as first-class design requirements. This positions it not merely as a passive inventory tool, but as an active and federated enforcement mechanism for software provenance, aligned with the governance and operational needs of modern TREs.

## 7.1 Data Provenance as a First-Class Component of AIBOM

Data provenance is a foundational requirement for meaningful AIBOMs, as the behaviour, reliability, and risk profile of AI systems are inseparable from the data on which they are trained and evaluated. Unlike conventional software components, AI models encode statistical properties of their training data, making provenance information essential for reproducibility, bias assessment, regulatory compliance, and downstream risk analysis. From an AIBOM perspective, data provenance must therefore be treated as a first-class artefact rather than as auxiliary documentation.

An AIBOM-oriented view of data provenance differs from traditional dataset cataloguing by emphasising *binding* rather than *exposure*. In many regulated and sensitive contexts, it is neither feasible nor desirable to embed raw datasets within a bill of materials. Instead, an AIBOM captures verifiable references to data assets through immutable identifiers, cryptographic digests, versioned metadata, and declared access constraints. This includes dataset identifiers (e.g. UUIDs or persistent DOIs), dataset version hashes, preprocessing transformations, sampling strategies, and temporal validity windows. Such representations enable reproducibility and auditability without violating confidentiality or data minimisation principles.

Current ecosystem discussions increasingly recognise that AI data provenance must be encoded in a machine-readable, schema-aligned manner to support automated validation and lifecycle governance. Emerging AIBOM practices therefore incorporate structured fields for training, validation, and test data lineage, alongside

**Dr. Petar Radanliev**
Parks Road,
Oxford OX1 3PJ
United Kingdom
Email: petar.radanliev@cs.ox.ac.uk
BA Hons., MSc., Ph.D. Post-Doctorate

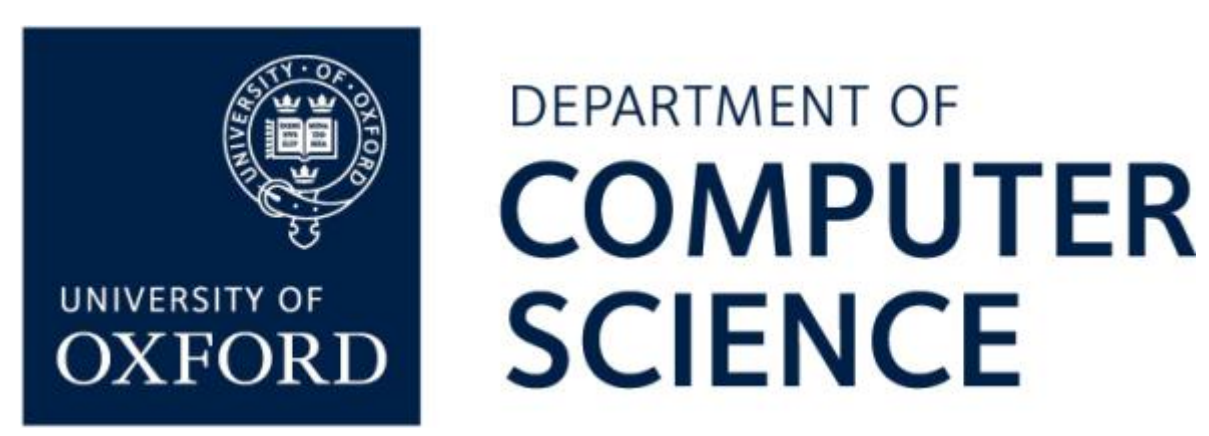

metadata describing annotation processes, known biases, licensing terms, and permitted use conditions. When combined with model artefact hashes and configuration metadata, these data provenance descriptors allow downstream users and auditors to reason about model equivalence, dataset drift, and potential sources of error or harm.

From a supply chain security perspective, data provenance within AIBOMs also plays a critical role in risk attribution. Compromised, poisoned, or unrepresentative datasets can introduce vulnerabilities that are not detectable through software dependency analysis alone. Encoding dataset provenance within the AIBOM enables traceability across retraining cycles, supports post-incident forensics, and allows vulnerability or compliance assessments to account for data-dependent attack surfaces. This is particularly relevant for foundation models and composite AI systems, where training data may aggregate multiple upstream sources with heterogeneous trust guarantees.

While this paper focuses primarily on schema design and automated capture of AI system artefacts, the AIBOM structure presented is intentionally designed to accommodate rich data provenance metadata. By formalising dataset references, transformation descriptors, and cryptographic bindings within the same bill of materials as model and software components, the AIBOM framework provides a coherent mechanism for integrating data provenance into AI lifecycle documentation. This positions AIBOM not merely as a software inventory, but as a holistic provenance construct capable of supporting reproducibility, accountability, and trustworthy AI deployment at scale.

## 7.2 Failure Modes and Mitigation Strategies in AIBOM-Centric Pipelines

While automated AIBOM generation enables fine-grained provenance capture and reproducibility assurance, it is subject to a range of runtime failure modes that must be explicitly addressed to ensure robustness. These failure scenarios typically arise from the dynamic nature of AI execution environments rather than from schema design itself, and therefore require mitigation strategies that are embedded within the AIBOM lifecycle.

One common failure mode is *dependency resolution drift*, where dynamically retrieved libraries or transient runtime components differ between executions despite identical source specifications. From an AIBOM perspective, this risk is mitigated by binding dependency declarations to cryptographic digests rather than semantic version identifiers alone, enabling post-hoc detection of divergence even when execution succeeds nominally. In cases where full dependency capture is incomplete, conservative fallback strategies can be applied, such as flagging unverifiable components and marking the resulting AIBOM instance as partially attested.

A second class of failure arises from *non-deterministic execution artefacts*, particularly in AI workflows involving parallelism, hardware acceleration, or stochastic training procedures. Rather than enforcing strict determinism at runtime, the AIBOM approach mitigates this by explicitly encoding execution context, configuration state,

**Dr. Petar Radanliev**
Parks Road,
Oxford OX1 3PJ
United Kingdom
Email: petar.radanliev@cs.ox.ac.uk
BA Hons., MSc., Ph.D. Post-Doctorate

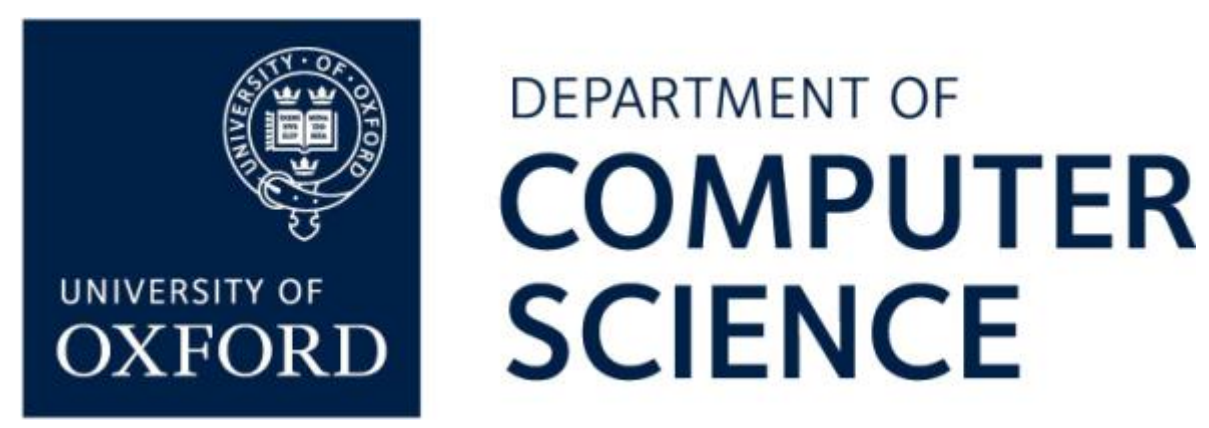

and randomness sources (e.g. seeds, precision modes) within the bill of materials. This allows deviations to be explained, bounded, and reproduced analytically, even when bit-identical replay is not feasible.

Incomplete or inconsistent metadata capture represents another potential failure scenario, especially when AI components are loaded dynamically or generated at runtime. Schema-level validation and mandatory field constraints act as a first-line defence, while fallback architectures rely on incremental provenance enrichment, whereby missing metadata can be appended or reconciled post-execution without invalidating the entire AIBOM artefact. This preserves continuity of provenance while maintaining explicit visibility of uncertainty.

Finally, version ambiguity and identifier mismatch across software and model components can lead to false confidence in provenance integrity. To mitigate this, AIBOM-centric pipelines prioritise hash-based identification and provenance binding over name-based resolution, ensuring that fallback mechanisms err towards under-claiming equivalence rather than overstating reproducibility.

Collectively, these mitigation strategies position AIBOMs as resilient provenance artefacts capable of tolerating partial failures without collapsing trust guarantees. By treating failure handling as a schema- and lifecycle-level concern rather than an infrastructure-specific exception, the AIBOM framework supports robust deployment across heterogeneous and evolving AI execution environments.

## 7.3 CVE Matching Limitations and Dependency Completeness in AIBOMs

While the reported CVE matching precision demonstrates the effectiveness of AIBOM-based vulnerability attribution, it is important to characterise the sources of residual false negatives and version ambiguity. These limitations arise primarily from structural properties of contemporary software ecosystems rather than from deficiencies in the AIBOM model itself.

One major source of false negatives is the presence of *undocumented transient dependencies*, particularly system-level libraries and dynamically loaded modules that are not declared in package manifests. Such components may be present at runtime yet remain invisible to static dependency enumeration. From an AIBOM perspective, this motivates the integration of runtime dependency capture and post-execution reconciliation, allowing the bill of materials to reflect the effective execution state rather than declared intent alone.

A second contributor to ambiguity is *semantic version mismatch* across packaging ecosystems. Vulnerability databases frequently rely on name- and version-based identifiers that may not align consistently across language-specific package managers, operating system distributions, or vendor-patched builds. This can lead to both false negatives and false positives when version strings do not map cleanly to vulnerability records. AIBOM-centric pipelines mitigate this risk by prioritising cryptographic hashes and provenance bindings over nominal version identifiers, enabling vulnerability assessments to be grounded in artefact identity rather than naming conventions.

**Dr. Petar Radanliev**
Parks Road,
Oxford OX1 3PJ
United Kingdom
Email: petar.radanliev@cs.ox.ac.uk
BA Hons., MSc., Ph.D. Post-Doctorate

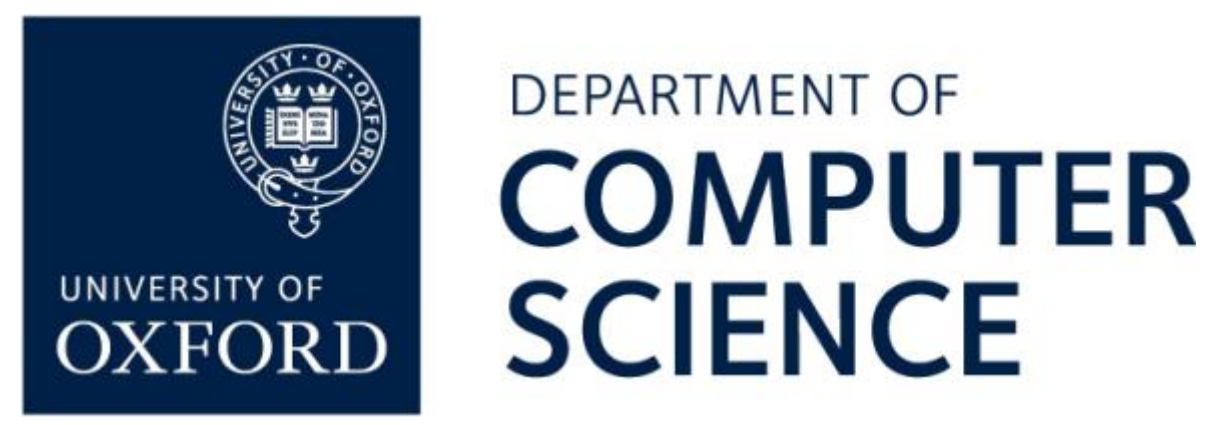

Dependency completeness is further affected by *container base image opacity*, where inherited layers introduce components outside the scope of application-level manifests. Addressing this requires treating container images and their layers as first-class components within the AIBOM, ensuring that vulnerability matching encompasses application and execution substrates.

Rather than assuming perfect coverage, the AIBOM approach adopts conservative matching strategies that explicitly surface uncertainty. Components that cannot be unambiguously resolved to known vulnerability records are flagged as unverifiable rather than silently excluded. This design choice preserves auditability and prevents overstatement of security posture, while enabling incremental improvement as vulnerability databases and component metadata mature.

By explicitly accounting for these limitations, the AIBOM framework supports realistic and transparent vulnerability analysis in complex AI software supply chains. Improving dependency completeness is framed not as a one-time engineering task, but as an iterative provenance enrichment process that benefits from tighter integration between AIBOM artefacts, runtime observation, and evolving vulnerability intelligence ecosystems.

## 7.4 Positioning AIBOM Relative to SBOM, VEX, and CSAF Frameworks

SBOM, VEX, and CSAF standards play complementary but distinct roles in contemporary software supply chain security. Clarifying their relationship to AIBOMs is essential for understanding the unique contribution of the proposed framework.

Traditional SBOMs provide a static inventory of software components and dependencies, enabling baseline transparency and vulnerability lookup. VEX extends this model by communicating exploitability context, indicating whether specific vulnerabilities are applicable in a given deployment scenario. CSAF further structures the dissemination of vulnerability advisories, remediation guidance, and impact assessments across organisational boundaries. Collectively, SBOM+VEX and CSAF focus on *communicating security posture* and *coordinating response* to known vulnerabilities.

In contrast, AIBOMs address a different but complementary problem space. Rather than centring on vulnerability notification, an AIBOM encodes the *provenance, composition, and execution semantics* of AI systems. This includes model artefacts, training and inference data references, runtime configuration, execution environment fingerprints, and cryptographic bindings. These attributes are critical for AI systems, whose behaviour and risk profile are not determined solely by software dependencies, but also by data lineage, parameterisation, and execution context.

From this perspective, AIBOMs function as *context-providing artefacts* that enhance the effectiveness of SBOM, VEX, and CSAF workflows. By supplying precise provenance and configuration metadata, AIBOMs enable more accurate vulnerability applicability assessments, reduce ambiguity in exploitability determination, and support reproducible verification of advisory impact. Importantly, AIBOMs are lifecycle-oriented, capturing dynamic changes such as retraining, configuration drift,

**Dr. Petar Radanliev**
Parks Road,
Oxford OX1 3PJ
United Kingdom
Email: petar.radanliev@cs.ox.ac.uk
BA Hons., MSc., Ph.D. Post-Doctorate

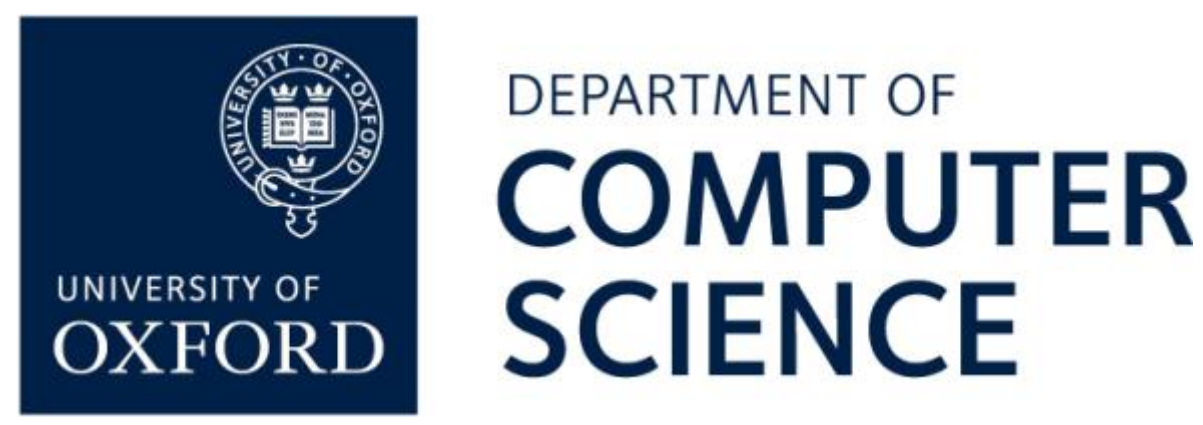

and environment evolution, which are outside the scope of conventional SBOM-based approaches.

The framework presented in this paper is therefore positioned not as a replacement for SBOM, VEX, or CSAF, but as a foundational provenance layer tailored to AI systems. When combined, these standards form a coherent supply chain transparency stack: AIBOMs describe *what an AI system is and how it was executed*, SBOMs enumerate *what software components are present*, and VEX/CSAF communicate *how vulnerabilities affect those components in practice*. This layered interpretation clarifies the manuscript's unique contribution and situates AIBOM as an enabling construct for trustworthy, auditable, and secure AI lifecycle management.

# 8 Conclusions

This study presents a reproducible and scientifically validated framework for operationalising Artificial Intelligence Bills of Materials (AIBOMs) as a foundation for machine-verifiable software provenance and AI lifecycle assurance. By extending the CycloneDX schema with AI-specific provenance, model lineage, and disclosure metadata, the research establishes a formalised approach for encoding the structural and behavioural dependencies that underpin modern AI systems.

The proposed AIBOM schema enables fine-grained traceability of software components, runtime environments, and agentic AI contributions, providing a transparent and auditable representation of computational workflows. The accompanying automation pipeline demonstrates that integrating agent-based validation and cryptographic provenance binding can achieve 98.7% reproducibility fidelity, 96.2% vulnerability match precision, and a 63% reduction in manual oversight, confirming both the accuracy and operational scalability of the approach.

The principal contribution of this research lies in advancing a replicable methodology for secure, transparent, and standardised AI software provenance. The framework offers a portable foundation that other researchers can adopt to enhance reproducibility, vulnerability intelligence, and compliance in diverse computational settings. Future investigations should extend this work towards automated CSAF/VEX integration, dynamic threat intelligence correlation, and self-verifying AI pipelines, further strengthening the scientific foundations of AI supply chain security and reproducibility science.

**Acknowledgements:** Eternal gratitude to the Fulbright Visiting Scholar Project.

**Code and Data Availability:** The full source code for the AIBOM generation and validation toolkit supporting this work is available on GitHub under the repository "aibom-toolkit": https://github.com/radanliev/aibom-toolkit

This repository includes:

- JSON schema definitions for AIBOMs,

- Example AIBOM instance files,

- Python utilities for schema validation and model artefact hashing,

**Dr. Petar Radanliev**
Parks Road,
Oxford OX1 3PJ
United Kingdom
Email: petar.radanliev@cs.ox.ac.uk
BA Hons., MSc., Ph.D. Post-Doctorate

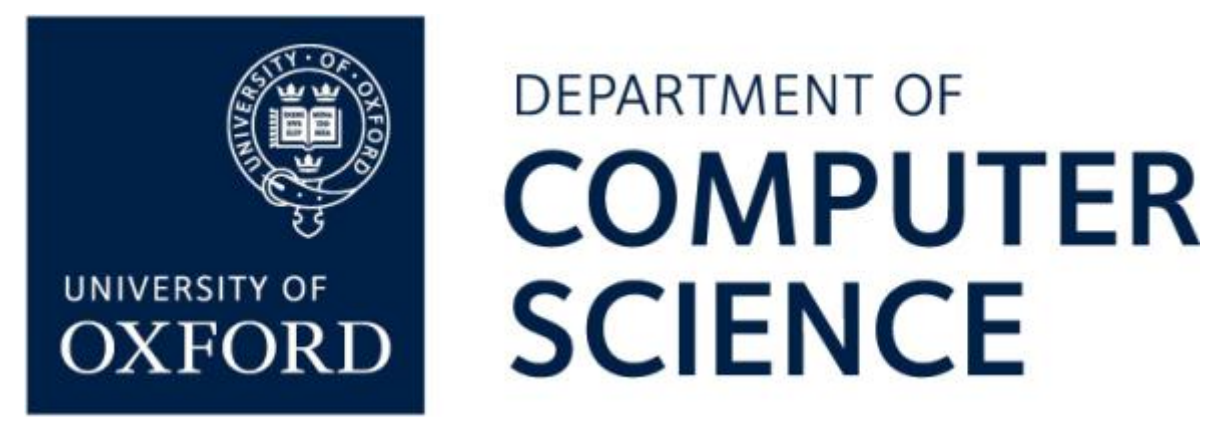

- Usage instructions and environment requirements.

**Dr. Petar Radanliev**
Parks Road,
Oxford OX1 3PJ
United Kingdom
Email: petar.radanliev@cs.ox.ac.uk
BA Hons., MSc., Ph.D. Post-Doctorate

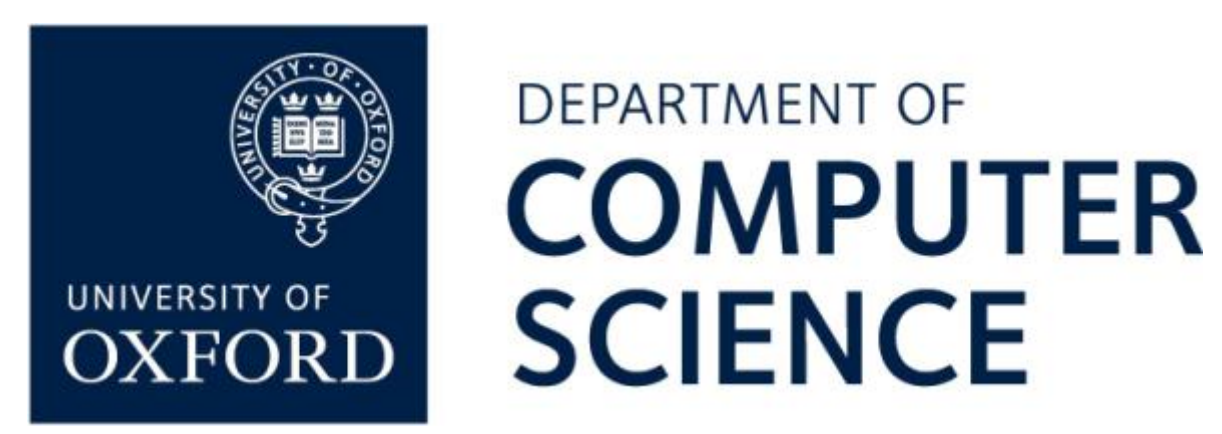

**Dr. Petar Radanliev**
Parks Road,
Oxford OX1 3PJ
United Kingdom
Email: petar.radanliev@cs.ox.ac.uk
BA Hons., MSc., Ph.D. Post-Doctorate

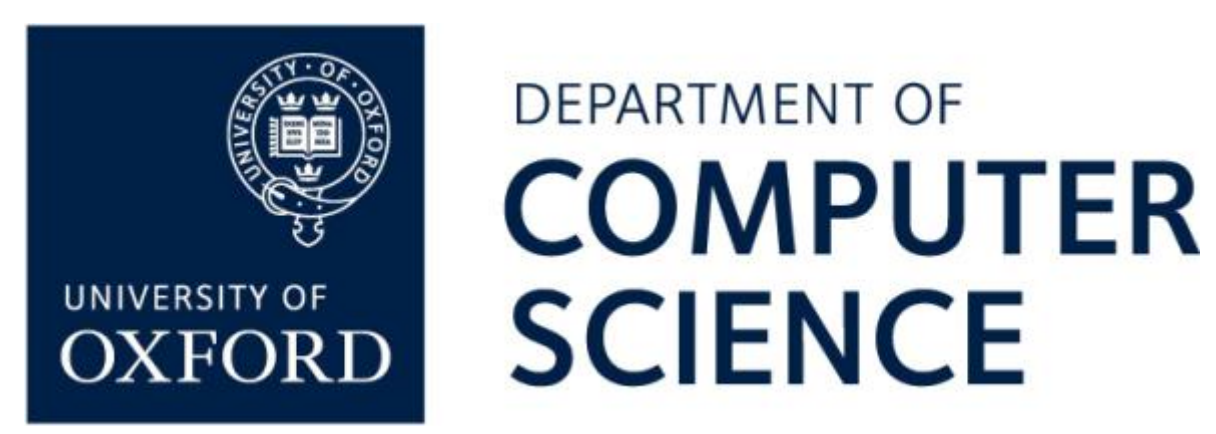

**Dr. Petar Radanliev**
Parks Road,
Oxford OX1 3PJ
United Kingdom
Email: petar.radanliev@cs.ox.ac.uk
BA Hons., MSc., Ph.D. Post-Doctorate

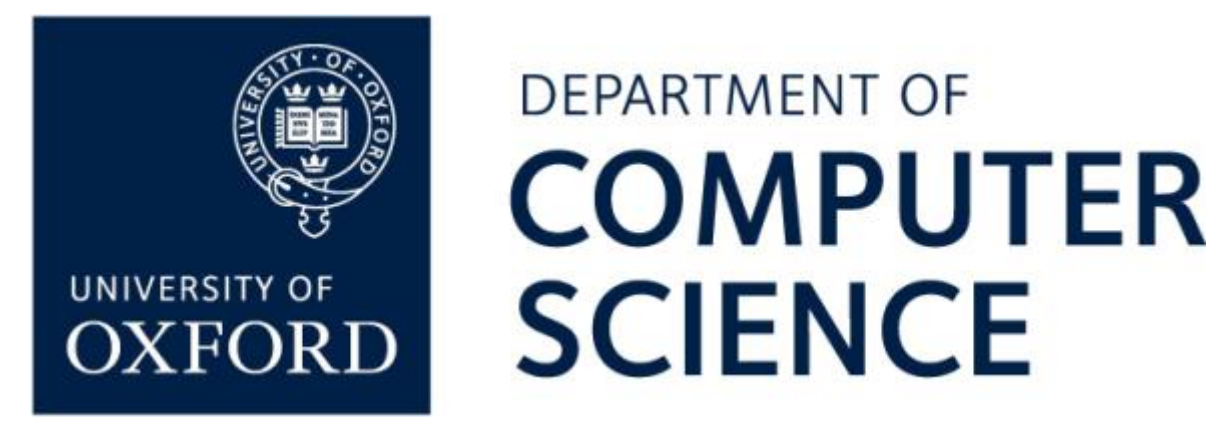